\documentclass[twocolumn,prb,10pt,aps,longbibliography,superscriptaddress]{revtex4-2}

\usepackage{graphicx}
\usepackage{amsmath}
\usepackage{amssymb}
\usepackage{amsfonts}
\usepackage{dsfont}
\usepackage{dcolumn}
\usepackage{bbm}
\usepackage{mathtools}
\usepackage{braket}
\usepackage{physics}
\usepackage{siunitx}
\usepackage[unicode]{hyperref}
\usepackage[nameinlink]{cleveref}
\Crefname{section}{Sec.}{Secs.}
\Crefname{equation}{Eq.}{Eqs.}
\Crefname{figure}{Fig.}{Figs.}
\Crefname{tabular}{Tab.}{Tabs.}

\usepackage{bbold} 
\usepackage{nicefrac}
\usepackage{tikz}
\usetikzlibrary{arrows.meta}
\usepackage{lipsum}  
\usepackage{xcolor}

\usepackage[makeroom]{cancel}

\definecolor{nat_green}{HTML}{43B02A}

\definecolor{tdo_green}{HTML}{83B818}
\definecolor{tdo_darkgreen}{HTML}{839A00}

\newcommand{\bes}{\begin{subequations}}
\newcommand{\ees}{\end{subequations}}
\newcommand{\be}{\begin{equation}}
\newcommand{\ee}{\end{equation}}

\newcommand{\sumnn}{\sum_{\langle i,j\rangle}}
\newcommand{\kk}{\mathbf{k}}
\newcommand{\lr}[1]{\left( #1 \right)}

\newcommand{\lrg}[1]{\left\{#1\right\}}
\newcommand{\lara}[1]{\langle #1 \rangle}

\newcommand{\nbi}[1]{#1_i^\dagger {#1}_i}
\newcommand{\nbk}[1]{#1_\mathbf{k}^\dagger {#1}_\mathbf{k}}
\newcommand{\nbj}[1]{#1_j^\dagger {#1}_j}
\newcommand{\hc}{\text{h.c.}}
\newcommand{\ha}{h^z_\text{alt}}
\newcommand{\sumk}{\sum_{\mathbf{k}}}

\newcommand{\ak}{a_\mathbf{k}}
\newcommand{\bk}{b_\mathbf{k}}
\newcommand{\akm}{a_\mathbf{-k}}
\newcommand{\bkm}{b_\mathbf{-k}}

\newcommand{\gam}{\gamma_\mathbf{k}}

\newcommand{\alk}{\alpha_\mathbf{k}}
\newcommand{\bek}{\beta_\mathbf{k}}
\newcommand{\alkm}{\alpha_\mathbf{-k}}
\newcommand{\bekm}{\beta_\mathbf{-k}}

\newcommand{\thetak}{\theta_\mathbf{k}}

\begin{document}
    \title{Switching the Magnetization in Quantum Antiferromagnets}

    \author{Katrin Bolsmann}
    \email{katrin.bolsmann@tu-dortmund.de}
    \affiliation{Condensed Matter Theory, 
    Technische Universit\"{a}t Dortmund, Otto-Hahn-Stra\ss{}e 4, 44221 Dortmund, Germany}

    \author{Asliddin Khudoyberdiev}
    \email{asliddin.khudoyberdiev@tu-dortmund.de}
    \affiliation{Condensed Matter Theory, 
    Technische Universit\"{a}t Dortmund, Otto-Hahn-Stra\ss{}e 4, 44221 Dortmund, Germany}

    \author{G\"otz S.\ Uhrig}
    \email{goetz.uhrig@tu-dortmund.de}
    \affiliation{Condensed Matter Theory, 
    Technische Universit\"{a}t Dortmund, Otto-Hahn-Stra\ss{}e 4, 44221 Dortmund, Germany}

    \date{\textrm{\today}}

    \begin{abstract}
    The orientation of the order parameter of quantum magnets can be used to store information
		in a dense and efficient way. Switching this order parameter corresponds to writing
		data. To understand how this can be done, we study a precessional reorientation 
		of the sublattice magnetization  in an (an)isotropic quantum antiferromagnet
		induced by an applied magnetic field. We use a description including the leading quantum and
		thermal fluctuations, namely Schwinger boson mean-field theory, because this theory allows us to
		describe both ordered phases and the phases in between them, as is crucial for switching.
		An activation energy has to be overcome requiring a minimum applied field $h_\text{t}$
		which is given essentially by the spin gap. It can be reduced significantly for temperatures approaching the
		N\'eel temperature facilitating switching. The time required for switching diverges when the
		field approaches $h_\text{t}$ which is the signature of an inertia in the magnetization dynamics.
		The temporal evolution of the magnetization and of the energy reveals signs of dephasing.
		The switched state has lost a part of its coherence because the magnetic modes 
		do not evolve in phase.
		\end{abstract}

    \maketitle

    \section{Introduction}
    \label{s:introduction}
		
Spintronics aims at exploiting the additional degree of freedom
represented by the electron spin. In the early days of this 
very active field quantum antiferromagnets did not play a big role
seemingly corroborating N\'eel's famous quote ``They are extremely 
interesting from the theoretical viewpoint, but do not seem to have any applications.''
\cite{neel70}. Indeed, ferromagnets have an advantage in their
ease of measurement and manipulation \cite{misra11}.
		
Contrary, however, to N\'eel's view the focus of spintronic research in
experiment and theory shifted towards antiferromagnets 
in the last years, as they have many advantages over ferromagnets \cite{gomon17}.
Ferromagnetic domains, for instance, exhibit stray fields, 
which affect neighboring domains. This is detrimental for data storage and switching
because adjacent bits influence one another.
Since the stray-field interactions become the stronger the closer the bits are packed, 
the maximum density of bits is limited.
But the era of big data and digitalization requires more and more 
storage capacity in smaller and smaller physical space \cite{chen14,hilbe11}.
Loth et al.\ demonstrated in 2012 that it is possible to substantially reduce the 
distance between antiferromagnetic bits compared to ferromagnetic bits
due to the absence of stray fields \cite{loth12}.

Another significant advantage of antiferromagnets is that their eigenfrequencies are in the 
THz range while the ferromagnetic ones lie in the GHz range \cite{gomon14,kampf11}.
Thus one may expect that the typical times for manipulations are also
shorter by a factor 1000 than in ferromagnets.
These and other advantageous properties are the reason why measurement 
and control of antiferromagnets have become one of the most important fields of research in 
spintronics over the past decade \cite{jungw16}. 

There are various methods to read out the 
direction of the N\'eel vector of an antiferromagnet, even though it has no macroscopic magnetization.
One way is by electrical measurements of the magnetoresistance \cite{baldr18,bodna20}, 
but there are also approaches based on optical means \cite{grigo21}.
The manipulation of antiferromagnetic order has also been realized by 
various techniques \cite{kimel09,cheng15,gomon10}. In 2016, for instance, 
based on the theoretical predictions of \v{Z}elezn\'y et al.\ \cite{zelez14}, 
Wadley et al.\ \cite{wadle16} succeeded in switching the N\'eel vector 
of the antiferromagnet CuMnAs using current-induced internal fields.

In this work, we present a microscopic description of a two-dimensional 
anisotropic quantum antiferromagnet in and out of equilibrium based on  spin-wave theory
which captures the leading quantum and thermal fluctuations. 
Our main goal is to describe precessional switching of the sublattice magnetization
by external magnetic fields.
In our approach not only the magnetization at the zero and at the staggering
wave vector is tracked, but \emph{all} magnonic modes at \emph{all} wave vectors. This includes the effects
of dephasing, i.e., the fact that the modes evolve with different frequencies.
This allows us to describe interference effects faithfully. Moreover, a description
based on the elementary magnons allows us also 
to capture the influence of finite temperature, at least on the mean-field level.
We emphasize that, to our knowledge, so far the theoretical approaches were based
on the two vectors of the average magnetizations on the two sublattices, see all references 
above. Thus, our work reports conceptional and methodological progress apt to 
improve our understanding of the switching dynamics in quantum antiferromagnets.

Of course, spin-wave theory is a standard tool \cite{auerb94}. But the most common
representations proposed by Holstein and Primakoff \cite{holst40} or by Dyson and Maleev
\cite{dyson56a,malee58} start from a particular ordered state. The magnons describe only small deviations
from one of the N\'eel states. In view of the goal to capture switching the 
magnetization from up-down on the two sublattices to down-up or vice versa
this is insufficient because the switching process includes the two 
antagonal N\'eel states. Starting from one of them and reaching its 
antipole implies far more than a small deviation. Hence, a suitable
bosonic representation is required for which we use Schwinger bosons 
\cite{schwi52,arova88,auerb88,auerb94,manue99}. 

This representation does not use long-range order in a N\'eel state as
reference state so that it can describe even the disordered
state. Thus, even on the mean-field level, Schwinger bosons can capture large deviations
from one of the ground states. We are aware that Schwinger boson mean-field theory 
may display pathologies in form of spurious first-order transitions.
But these are not likely in unfrustrated low-dimensional systems \cite{silva02}.
We use Schwinger bosons to describe the isotropic and anisotropic systems at equilibrium initially. 
Then, we demonstrate how this representation can be used to simulate the switching 
of the sublattice magnetization by means of an external magnetic field.
We analyze how the system changes after the rotation and how the anisotropy influences
the switching. Thereby our study provides a basis for further theoretical investigations 
regarding the full control of quantum antiferromagnets which in turn is 
expected to guide further experimental investigations.

The article is set up as follows. After this Introduction, Sect.\ \ref{s:model}
introduces the model and its bosonic representation by Schwinger bosons.
Then, Sect.\ \ref{s:equilibrium}
briefly recalls the corresponding mean-field theory in equilibrium for the isotropic
and the anisotropic Heisenberg antiferromagnet. Subsequently, 
we derive the equations of motion describing the dynamics in applied magnetic
fields which induce precessional motion in Sect.\ \ref{s:switch}. 
Finally, we summarize our results in the Conclusions \ref{s:conclusion}.

    \section{Heisenberg model and its Schwinger boson representation}
    \label{s:model}

\subsection{Heisenberg model and its Hamiltonoperator}
\label{ss:model}

We consider here the (an)isotropic Heisenberg model on a square lattice for
a quantum antiferromagnet with localized spins $S=1/2$. Its Hamilton operator reads
\begin{align}
  \mathcal{H} &= \sumnn \left\{ J_{xy}\lr{S^x_iS^x_j + S^y_iS^y_j} 
	+ J_z S^z_i S^z_j \right\}  ,
	\label{eqn:aniAFM-H2}
\end{align} 
where $i$ and $j$ label sites of the underlying lattice, the sum runs over pairs of nearest neighbors
counting each pair only once. The operators	$S^\alpha_i$ are the usual operators of the spin component
$\alpha$ at site $i$; the couplings $J_z$ and $J_{xy}$ are both antiferromagnetic, i.e., positive.
We focus on the easy-axis model taking $J_z=J$ as energy unit and defining the ratio 
$\chi=J_{xy}/J_z\in [0,1]$ so that the Hamiltonian can be rewritten to
\begin{subequations}
    \begin{align}
      \mathcal{H} &= J \sumnn \left\{\chi\lr{S^x_iS^x_j + S^y_iS^y_j} + S^z_i S^z_j \right\} \\
      &= J \sumnn \left\{\frac{\chi}{2}\lr{S^+_iS^-_j + S^-_iS^+_j} + S^z_i S^z_j \right\}
			,\label{eqn:aniAFM-H}
    \end{align}
\end{subequations}
where the spin ladder operators are used in the last step.

For finite anisotropy $\chi<1$, the magnons display a finite energy gap $\Delta(\chi)>0$. 
In turn, the spin-spin correlation length $\xi$ becomes finite following the estimate
\begin{equation}\label{eqn:corrlen}
    \xi(\chi) = \frac{v}{\Delta(\chi)}, 
\end{equation}
where $v$ is the spin-wave velocity. Below, we will perform a calculation for finite
clusters with linear dimension $L$. These will reflect the thermodynamic, infinite size
results whenever $L\gg\xi$ holds.

A uniform magnetic field can be included by adding the term
\begin{equation}\label{eqn:uniform}
    \mathcal{H}_{\text{uni}} = - \mathbf{h}\cdot\sum_i \mathbf{S}_i ,
\end{equation}
while an alternating (staggered) magnetic field is accounted for by adding
\begin{equation}\label{eq:alternat}
    \mathcal{H}_{\text{alt}} = - \mathbf{h}_\text{alt}\cdot\sum_i (-1)^i \mathbf{S}_i.
\end{equation}
The length of the vector $\mathbf{h}$ is given as usual by $g\mu_\text{B}B$.

\subsection{Schwinger boson representation}
\label{ss:schwinger}

Introduced by Schwinger in 1952 \cite{schwi52}, 
this representation uses two boson flavors $a_i^{(\dag)}$ and $b_i^{(\dag)}$ at site $i$
to transform the spin operators as follows
\bes
\label{eqn:SB-AFM}
\begin{align}
        S_i^+ &= a_i^\dagger b_i,
\label{eq:s+}
\\
        S_i^- &= b_i^\dagger a_i,
\label{eq:s-}
\\
        S_i^z &= \frac{1}{2}\lr{\nbi{a} - \nbi{b}},
\end{align}
\ees
so that the spin commutations relations are reproduced.
In contrast to the Holstein-Primakoff \cite{holst40} and Dyson-Maleev bosons
\cite{dyson56a,malee58}, both flavors of Schwinger bosons act on any lattice site 
and are not restricted to one sublattice. The physical meaning of the bosons can be explained best by taking a closer 
look at the expectation value 
\begin{equation}
\label{eq:magnetization}
    \langle S_i^z \rangle = \frac{1}{2}\lr{\langle \nbi{a} \rangle  - \langle \nbi{b}\rangle}.
\end{equation}
This corresponds to the difference between the mean occupation of the $a$-bosons and $b$-bosons at lattice site $i$.
The expectation value $\langle S_i^z \rangle$ is maximized if there are only $a$-bosons at that lattice site 
and it is minimized if there are only $b$-bosons. Expectation values lying between these two extremes 
are obtained by a mixture or a superposition of the two boson flavors. Given the finite value of the spin length $S$,
 it is obvious that the number of bosons per lattice site cannot be arbitrary since the expectation 
value of a spin along one axis can never be larger than $S$: $|\langle S_i^z \rangle| \leq S$. 
To guarantee that only the physical subspace is considered, $\mathbf{S}_i^2 = S(S+1)$ must hold.
This condition is equivalent to a local constraint on the number of bosons per lattice site, namely
\begin{equation}\label{eqn:SchwingerConstr}
    \nbi{a} + \nbi{b} = 2S \qquad \forall\, i \in \mathbb{N} \leq N  .
\end{equation}
In  the mean-field approach, we refrain from fulfilling this constraint at each site.
Instead, we include the constraint in the Hamiltonian as a Lagrange multiplier
to ensure \eqref{eqn:SchwingerConstr} on average.
By creating a boson of one flavor and annihilating a boson of the other flavor, the ladder operators $S^\pm$ 
in Eqs.\ \eqref{eq:s+} and \eqref{eq:s-} realize 
transitions  between the eigenstates of different magnetic quantum numbers.

Using the Schwinger representation \eqref{eqn:SB-AFM}, 
the anisotropic antiferromagnetic Hamiltonian~\eqref{eqn:aniAFM-H} reads
\begin{subequations}
    \begin{align} \nonumber
        \mathcal{H} &= J \sumnn \left\{\frac{\chi}{2}\lr{a_i^\dagger b_i  b_j^\dagger a_j 
				+ a_j^\dagger b_j b_i^\dagger a_i}\right.\\
				& \quad + \left. \frac{1}{4}\lr{\nbi{a} - \nbi{b}}\lr{\nbj{a} - \nbj{b}} \right\}  
				\\ \nonumber
			  & = J \sumnn \Big\{\frac{\chi}{2}\lr{a_i^\dagger b_i  b_j^\dagger a_j
				+ a_j^\dagger b_j b_i^\dagger a_i} +\\ 
         & \quad\frac{1}{4}\lr{\nbi{a}\nbj{a} - \nbi{a}\nbj{b} 
				- \nbi{b}\nbj{a} + \nbi{b}\nbj{b}} \Big\} .
							\label{eqn:SchwingerH}
        \end{align}
\end{subequations}
The model can be extended to $M$ flavors of the Schwinger bosons
allowing for an  SU($M$) symmetry \cite{auerb94,rayki93,mathu10}.
In this work, however, we stick to two boson flavors $a$ and $b$ since 
we deal with a spin Hamiltonian with SU($2$)-symmetry.

    \section{Mean-field theory of the equilibrium}
    \label{s:equilibrium}
		
		Here we review the equilibrium solutions for the isotropic and the anisotropic
		cases for three reasons: (i) to introduce the notation, (ii) to discuss the subtleties
		of the Schwinger boson description, and (iii) to provide the initial conditions for the 
		magnetization switching considered in the following section.

	    \subsection{Spin isotropic case}
				\label{ss:isotropic}	
		
By setting $\chi=1$, the isotropic Schwinger Hamiltonian is obtained from \eqref{eqn:SchwingerH}
\begin{equation}
\begin{aligned}
    \mathcal{H} &= \sumnn \Big\{\frac{1}{2}\lr{a_i^\dagger b_i  b_j^\dagger a_j + a_j^\dagger b_j b_i^\dagger a_i}+ \\ 
         &\quad \frac{1}{4}\lr{\nbi{a}\nbj{a} - \nbi{a}\nbj{b} - \nbi{b}\nbj{a} + \nbi{b}\nbj{b}} \Big\} .
\end{aligned}
\end{equation}
We use the coupling constant $J$ as the energy unit so that we can set it to unity.
Defining the antiferromagnetic bond operator $A_{ij} \coloneqq a_ib_j - a_jb_i$, the Hamiltonian can be rewritten to 
\begin{equation}\label{eqn:SchwingerH2}
        \mathcal{H} = -\frac{1}{2}\sumnn \lr{A_{ij}^\dagger A_{ij}-2S^2},
\end{equation}
where the constraint \eqref{eqn:SchwingerConstr} was used.
It is useful to rotate one sublattice by $180^\circ$ to obtain a uniform description 
of all lattice sites with full translational invariance. When rotating around the $S^y$-axis, the 
$x$- and $z$-components of the spin operators change their signs. Consequently, in terms of Schwinger bosons, the substitution 
\begin{equation}\label{eqn:rot}
    a_j \to -b_j\,, \qquad \qquad b_j\to a_j
\end{equation}
is applied to one sublattice. The bond operator now reads
\begin{equation}
    A_{ij} = a_i a_j + b_ib_j 
\end{equation}
while the representation of the Hamiltonian~\eqref{eqn:SchwingerH2}
remains unaltered. Defining the expectation value 
\be 
\label{eq:A}
A \coloneqq \langle A_{ij}\rangle = \langle A_{ij}^\dagger \rangle
\ee
and replacing the quadrilinear terms by the terms with one or two
contractions according to Wick's theorem the Hamiltonian \eqref{eqn:SchwingerH2} 
is converted to the mean-field Hamiltonian
\begin{align}\label{eqn:SchwingerH3}
    \mathcal{H} &= E_\text{MF} + \lambda \sum_i \lr{\nbi{a} + \nbi{b}}-
		\nonumber
		\\
		&\qquad \frac{1}{2}A\sumnn \lr{a_ia_j + b_ib_j + \hc}
\end{align}
with the mean-field energy $E_\text{MF} \coloneqq N(A^2 +2S^2)$.
Furthermore, to ensure that the constraint~\eqref{eqn:SchwingerConstr} is always fulfilled on average, 
an additional sum with the Lagrange multiplier $\lambda$ is added.

Calculating the Fourier transformation of the Hamiltonian is the first step towards diagonalization
yielding
\begin{align}
\nonumber
\mathcal{H} &= E_\text{MF} + \sumk\Big\{\lambda \lr{\nbk{a} + \nbk{b}} 
-\\
& \qquad A\,\gam\lr{\ak\akm + \bk\bkm +  \hc}\Big\}
\label{eqn:SchwingerH4}
\end{align}
with $\gam \coloneqq \tfrac{1}{2}\lr{\cos(k_x) + \cos(k_y)}$ where we
set the lattice constant to unity.
Note that the sum refers to the entire Brillouin zone since the Schwinger bosons 
are not restricted to one sublattice. Next, standard Bogoliubov transformations 
for both the $a$- and $b$-bosons 
\bes
\begin{align}
        \ak^\dagger &= \cosh(\thetak)\alk^\dagger + \sinh(\thetak)\alkm,
				\\
				\bk^\dagger &= \cosh(\thetak)\bek^\dagger + \sinh(\thetak)\bekm,
\end{align}
\ees
and the corresponding Hermitian conjugate relations lead to the diagonalized Hamiltonian
\begin{equation}
\mathcal{H} = E_\text{MF} - N\lambda + \sumk \omega^\text{iso}_\mathbf{k}\lr{\nbk{\alpha} + \nbk{\beta} +1}, 
\end{equation}
where we have chosen $\lambda \tanh(2\thetak) = 2A\gam$. The resulting dispersion 
reads
\be
\label{SchwingerDispersion1}
\omega^\text{iso}_\mathbf{k} = \sqrt{\lambda^2-\lr{2A\gam}^2}.
\ee

The value $\lambda$ results from the condition that \eqref{eqn:SchwingerConstr} holds on average, i.e., for
the expectation values of the particle numbers. The expectation value $A$ is defined in \eqref{eq:A}
yielding 
\begin{subequations}
\label{eq:selfcon1}
\begin{align}
    A &= \frac{1}{N}\sumk \frac{2A\gam^2}{\omega^\text{iso}_\mathbf{k}}\coth(\beta\omega^\text{iso}_\mathbf{k}/2) , \label{eqn:SC1} \\
    2S + 1 &= \frac{1}{N}\sumk \frac{\lambda}{\omega^\text{iso}_\mathbf{k}}\coth(\beta\omega^\text{iso}_\mathbf{k}/2) . \label{eqn:SC2}
\end{align}
\end{subequations}
Here, $\beta$ is the inverse temperature up to Boltzmann's constant. Solving these equations requires 
finding a non-linear zero depending on two variables.
For finite systems $N<\infty$ or for finite temperature it can be tackled by direct numerics. 
No finite sublattice magnetization occurs because no finite system displays long-range antiferromagnetic order.
The same holds true for the infinite two-dimensional isotropic system at finite temperature 
according to the Mermin-Wagner theorem \cite{mermi66}.

But we briefly discuss the subtle occurrence of long-range order in the infinite system at zero
temperature in accordance with previous treatments \cite{arova88,auerb88,auerb94}. 
The Goldstone theorem \cite{auerb94}
tells us that in case of long-range order, the spectrum must be gapless, thus $\lambda=A$
holds. But then we encounter singularities at $\mathbf{k}=\mathbf{0}$ and $\mathbf{k}=(\mathbf{\pi,\pi})$
for finite $N$. In the thermodynamic limit $N\to\infty$, the sums in \eqref{eq:selfcon1} do not converge uniformly 
to integrals. The solution to this issue lies in the fact that for any finite $N$ the spectrum
is \emph{not} gapless, but displays a small finite-size gap $\Delta_N$. Although this gap vanishes
for $N\to\infty$ a contribution from the points $\mathbf{k}=\mathbf{0}$ and $\mathbf{k}=(\mathbf{\pi,\pi})$
remains. Concretely, we set $\lambda^2 = (2A)^2(1+\kappa^2)$ with $\kappa = {f/N}$
implying $\omega^\text{iso}_\mathbf{k} = 2A\sqrt{1 + \kappa^2 - \gam^2}$. Then the limit
$N\to\infty$ is performed for the Eqs.\ \eqref{eq:selfcon1} and we obtain
\begin{subequations}
\label{eq:selfcon2}
    \begin{align}
        A &= \frac{1}{4\pi^2}\int_\text{BZ} dk^2 \frac{\gam^2}{\sqrt{1 - \gam^2}} + \frac{2}{f} ,
				\\
	\label{eq:selfcon2b}			
        2S + 1 &= \frac{1}{4\pi^2}\int_\text{BZ} dk^2 \frac{1}{\sqrt{1 - \gam^2}} + \frac{2}{f}.
\end{align}
\end{subequations}
This contribution from single points in the Brillouin zone stands for the macroscopic contribution
of a few modes, here precisely four modes ($\alk, \bek$ at  $\mathbf{k}=\mathbf{0}$ and $(\mathbf{\pi,\pi})$),
representing a Bose-Einstein condensation. It is directly linked to the occurrence of long-range order because
one can show, see Appendix \ref{a:magnetization}, that the sublattice magnetization per site 
$m_0=|\langle S^z_i\rangle|$ in the ordered phase is given by
\bes
\begin{align}
\label{eq:magnet-condensate}
m_0&=1/f
\\
 &= S + \frac{1}{2} -\frac{1}{8\pi^2} \int_\text{BZ} d k^2 \frac{1}{\sqrt{1-\gam^2}} .
\label{eq:m0-explicit}
\end{align}
\ees
Subtracting the two equations \eqref{eq:selfcon2} yields
\bes
\begin{align}
2S + 1 - A &= \frac{1}{4\pi^2} \int_\text{BZ}  dk^2 \sqrt{1-\gam^2} = :2\delta + 1
\\
\Leftrightarrow\; A &= 2(S-\delta), \\
2\delta &= -0.15795,
\\
\omega^\text{iso}_\mathbf{k} &= 4(S-\delta)\sqrt{1 - \gam^2},
\end{align}
\ees
where we determined $2\delta$ numerically.

            \begin{figure}[htb]
                \centering
               \includegraphics[width=\columnwidth]{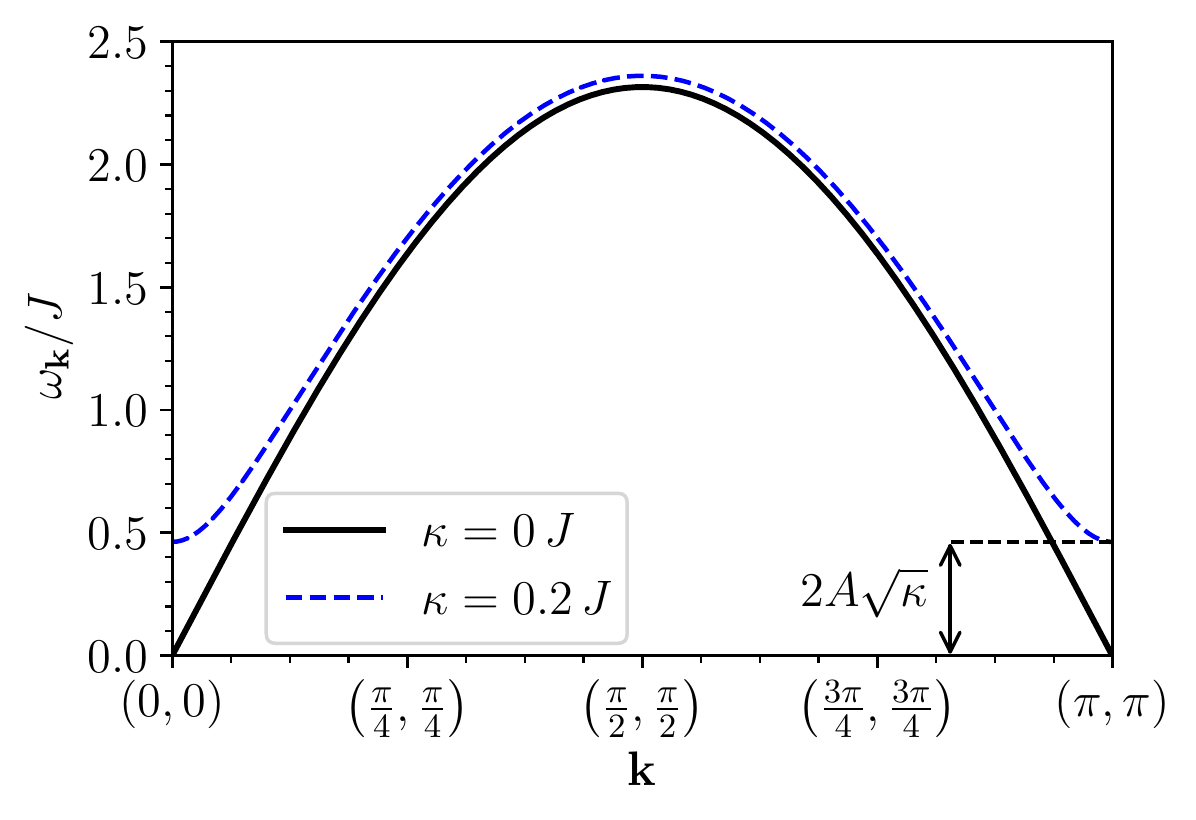}
                \caption{The black line depicts the degenerate dispersion of the
								two Schwinger bosons in the long-range ordered isotropic case in the thermodynamic
								limit. The blue dashed line illustrates the dispersion assuming a finite $\kappa=0.2\,J$.
								}
                \label{fig:isotrop-dispers}
            \end{figure}					

In Fig.\ \ref{fig:isotrop-dispers}, we show the resulting dispersion in the thermodynamic limit
(black line) and for an arbitrary finite $\kappa=0.2\,J$ (blue dashed line) for illustration. Note that
the black line shows the dispersion of both the $\alpha$- and the $\beta$-bosons because they
are degenerate. Since the spin operators consist of bilinear bosonic expressions, see Eq.\ 
\eqref{eqn:SB-AFM}, a physical excitation, i.e., the magnon, will not  
consist of a single $\alpha$ or $\beta$ particle. Given, however, the macroscopic occupations
at $\mathbf{k}=\mathbf{0}$ and $(\mathbf{\pi,\pi})$ it is justified to replace the bosonic
operators at these wave vectors according to
\bes
\begin{align}
\alk &\to \langle \alk\rangle = \sqrt{N m_0/4},
\\
\bek &\to \langle \bek\rangle = \sqrt{N m_0/4}
\end{align}
\ees
if no spontaneous symmetry breaking is accounted for. In the ordered phase,
only one of the bosons displays the macroscopic occupation, see Appendix \ref{a:magnetization}, so that
either $\alk \to  \sqrt{N m_0/2}$ or $\bek \to  \sqrt{N m_0/2}$ holds  at 
$\mathbf{k}=\mathbf{0}$ and $(\mathbf{\pi,\pi})$, see Ref.\ \cite{auerb94}.
With these substitutions the creation of a magnon is described by
$\sqrt{N m_0/2} \bek^\dag$ if the $\alpha$-bosons condense and 
by $\sqrt{N m_0/2} \alk^\dag$ if the $\beta$ bosons condense.
It is interesting to note that the dispersion from the Schwinger boson mean-field
theory is identical to the self-consistent spin-wave theory resulting from 
a $1/S$ expansion in the Holstein-Primakoff and in the Dyson-Maleev representation
up to and including order $1/S$.

Later we compute solutions for the self-consistency conditions numerically for finite
clusters with $N<\infty$ because we need them as initial conditions for switching processes. Rigorously,
no spontaneous symmetry breaking takes place due to the finiteness of the sample size
and thus no finite sublattice magnetization can be determined.
As a remedy to approximate the thermodynamic, infinite lattice, we include
a tiny symmetry-breaking alternating field $h_\text{alt}^z$. It turns out that a scaling
of this auxiliary field has to be chosen  $\propto 1/N$.
Then it reproduces the analytically known sublattice magnetization $m_0=S-0.19660$.
The details of the calculation are given in Appendices \ref{a:magnetization} and \ref{a:h-aux}.

	    \subsection{Spin anisotropic case}
				\label{ss:anisotropic}	
		
Here we address the anisotropic case and its description by a Schwinger boson mean field theory.
This case is more involved since the equations are less symmetric which requires to account 
for additional mean fields, see for instance Ref.\ \cite{ghiol15}. Conceptually, however,
there is an explicit difference in the anisotropic case between the two bosons so that no infinitesimal
fields are needed.

By using the constraint \eqref{eqn:SchwingerConstr}, the anisotropic Hamiltonian 
of the antiferromagnetic spin lattice \eqref{eqn:SchwingerH} can be written as
\begin{equation}\label{eqn:H_aniso}
    \mathcal{H} = -\frac{1}{4}\sumnn\!\!\!\lrg{(1+\chi)A_{ij}^\dagger A_{ij} + (1-\chi)B_{ij}^\dagger B_{ij} - 4S^2}
\end{equation}
with the bond operators $A_{ij} \coloneqq a_ia_j + b_ib_j$ and $B_{ij} \coloneqq a_ia_j - b_ib_j$.
As before, one sublattice has been rotated by $180^\circ$ using the substitution \eqref{eqn:rot}
so that all spins point in the same direction in the ordered phase.
The mean-field Hamiltonian is obtained by applying Wick's theorem neglecting quadrilinear normal-ordered
terms yielding
\begin{subequations}
\begin{align}
\nonumber
    \mathcal{H} = & E_\text{MF} + \lambda \sum_i \lr{\nbi{a} + \nbi{b}}-\\
		\nonumber
    & \quad \frac{1}{4}\sumnn\Big\{A(1+\chi)(a_ia_j + b_ib_j + \hc) +\\
    &\quad B(1-\chi)(a_ia_j - b_ib_j + \hc)\Big\} , 
		\label{eqn:SchwingerH6}
\end{align}
\end{subequations}
with $E_\text{MF} \coloneqq \frac{1}{2}N\lrg{A^2(1+\chi) + B^2(1-\chi) + 4S^2}$ and
 the expectation values $A \coloneqq \langle A_{ij}\rangle = \langle A_{ij}^\dagger\rangle$ 
and $B\coloneqq \langle B_{ij}\rangle = \langle B_{ij}^\dagger\rangle$.

Furthermore, a boson number term with the Lagrange multiplier $\lambda$ is added 
to ensure that the constraint~\eqref{eqn:SchwingerConstr} is also satisfied in the 
anisotropic case on average. With vanishing anisotropy $\chi\to 1$, this Hamiltonian 
corresponds to that of the isotropic system \eqref{eqn:SchwingerH3}.

In order to diagonalize the anisotropic Hamiltonian, we proceed as in the isotropic case.
After  Fourier transformation, the  Hamiltonian is given by
\begin{align}
\nonumber
\mathcal{H} & = E_\text{MF} + \lambda \sumk \lr{\nbk{a} + \nbk{b}}\\
\nonumber
    &\quad -\frac{1}{2}\sumk\gam\Big\{A(1+\chi)(\ak\akm + \bk\bkm + \hc) \\
    & \quad +B(1-\chi)(\ak\akm - \bk\bkm + \hc)\Big\}.
\end{align}
Since the non-diagonal Bogoliubov terms of the two Schwinger bosons flavors
have different prefactors, it is no longer possible to use the same 
Bogoliubov angles in the transformations. Thus we use
\bes
\begin{align}
  \ak^\dagger &= \cosh(\thetak^a)\alk^\dagger + \sinh(\thetak^a)\alkm,
	\\ \bk^\dagger &= \cosh(\thetak^b)\bek^\dagger + \sinh(\thetak^b)\bekm ,
\end{align} 
\ees
and the Hermitian conjugate operators 
using the condition 
\be
\gam = \frac{\lambda}{C_-}\tanh(2\thetak^a) = \frac{\lambda}{C_+}\tanh(2\thetak^b) 
\ee
with
\be
C_\pm \coloneqq\big(A\lr{1+\chi}\mp B\lr{1-\chi}\big).
			\label{eq:C-def}
\ee
In this way, we obtain the diagonal form
\begin{align}
    \mathcal{H} = E_\text{MF} - N\lambda + \sumk\{\omega^{-}_\mathbf{k}(\nbk{\alpha}+\frac{1}{2})+
		\omega^{+}_\mathbf{k}(\nbk{\beta}+\frac{1}{2})\} .
\end{align}
Concomitantly, there are two different spin-wave dispersions 
\be
    \omega^{\pm}_\mathbf{k} = \sqrt{\lambda^2-C_\pm^2\gam^2}.
\ee
For $\chi = 1$, the two dispersions coincide and reproduce
the isotropic dispersion~\eqref{SchwingerDispersion1}.
 
Physically, a spin gap is expected to  result from the anisotropy 
reducing the continuous symmetry to a discrete $\mathbb{Z}_2$-symmetry.
Therefore, spontaneous symmetry breaking due to the magnetic order 
no longer breaks a continuous symmetry and no massless Goldstone bosons need to occur.
Still, the Schwinger boson description of the ordered phase requires a
condensation of one Schwinger boson flavor at zero temperature.
Let us assume that the $\alpha$-bosons condense so that their dispersion is
gapless in the thermodynamic limit at $T=0$. We stress that this does not
imply that the physical excitations are gapless because they imply
the annihilation of an $\alpha$-boson combined with the creation of a
$\beta$-boson. 

\begin{figure}[htb]
    \centering
    \includegraphics[width=\columnwidth]{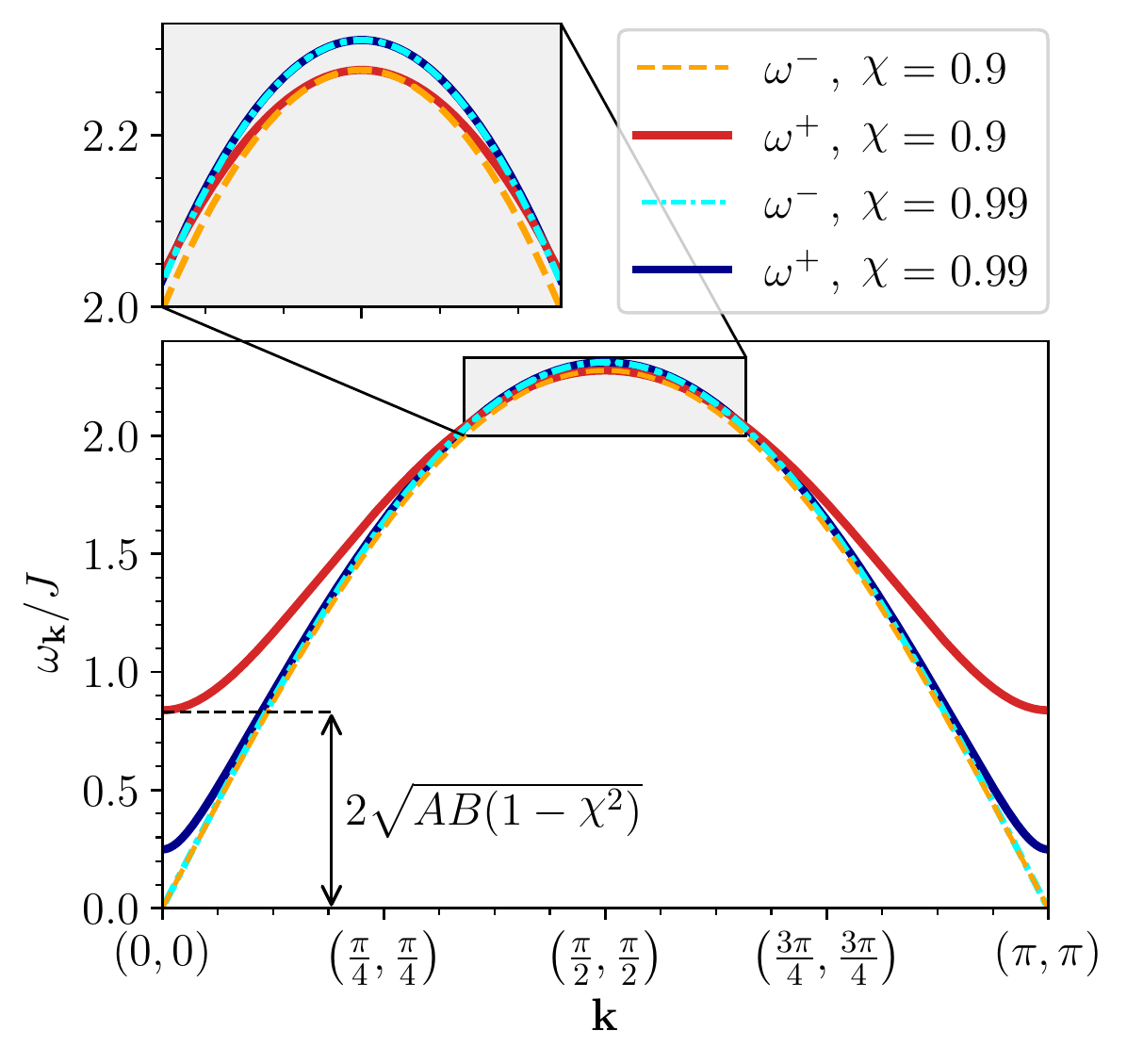}
    \caption{Dispersions $\omega^{\pm}_\mathbf{k}$ in an antiferromagnetic spin-$\frac{1}{2}$ square lattice 
		at zero temperature, plotted exemplarily for the two anisotropies $\chi = 0.9$ and $\chi = 0.99$. 
    For the same $\chi$, the maxima of the dispersions $\omega^{+}_\mathbf{k}$ 
    and $\omega^{-}_\mathbf{k}$ coincide, while $\omega^{-}_\mathbf{k}$ is gapless and
    $\omega^{+}_\mathbf{k}$ displays the physical energy gap at $\mathbf{k} = \lr{0,0}$ and $\mathbf{k} =
		\lr{\pi,\pi}$. Note that $\omega^{-}_\mathbf{k}$ does not describe observable modes.
    As expected, the energy gap $\Delta = 2\sqrt{AB(1-\chi^2)}$ increases with increasing
    anisotropy, i.e., increasing deviation of $\chi$ from 1.		}
    \label{fig:AnisoDisp}
\end{figure}

After these qualitative considerations, we need to derive 
the conditions for the parameters $A$, $B$, and $\lambda$ 
in order to be able to make any quantitative statements.
The constraint and the expectation  values $A$ and $B$  yield
\begin{subequations}
\label{eqn:SCEQ}
\begin{align}
    2S &= \frac{\lambda}{2N}\sumk\Bigg[\frac{\coth(\frac{\beta\omega^{-}_\mathbf{k}}{2})}{\omega^{-}_\mathbf{k}}
		+\frac{\coth(\frac{\beta\omega^{+}_\mathbf{k}}{2})}{\omega^{+}_\mathbf{k}}\Bigg] -1
		,\label{eqn:SCEQ1}\\
    A &= \frac{1}{2N}\sumk\gam^2\Bigg[\frac{C_-\coth(\frac{\beta\omega^{-}_\mathbf{k}}{2})}{\omega^{-}_\mathbf{k}}+
		\frac{C_+\coth(\frac{\beta\omega^{+}_\mathbf{k}}{2})}{\omega^{+}_\mathbf{k}}\Bigg]
		,\label{eqn:SCEQ2}\\
    B &= \frac{1}{2N}\sumk\gam^2\Bigg[\frac{C_-\coth(\frac{\beta\omega^{-}_\mathbf{k}}{2})}{\omega^{-}_\mathbf{k}}-
		\frac{C_+\coth(\frac{\beta\omega^{+}_\mathbf{k}}{2})}{\omega^{+}_\mathbf{k}}\Bigg].\label{eqn:SCEQ3}
\end{align}
\end{subequations}
These equations allow us to determine the dispersion at zero and finite
temperature. The case of zero temperature is again a bit subtle; the relevant treatment
of the equations at $T=0$ is described in Appendix \ref{a:magnetization-anisotropic}.
The resulting dispersion is displayed in Fig.\ \ref{fig:AnisoDisp}.
Clearly the spin gap in the physical dispersion appears and grows for larger
anisotropy.

\begin{figure}[htb]
    \centering
    \includegraphics[width=\columnwidth]{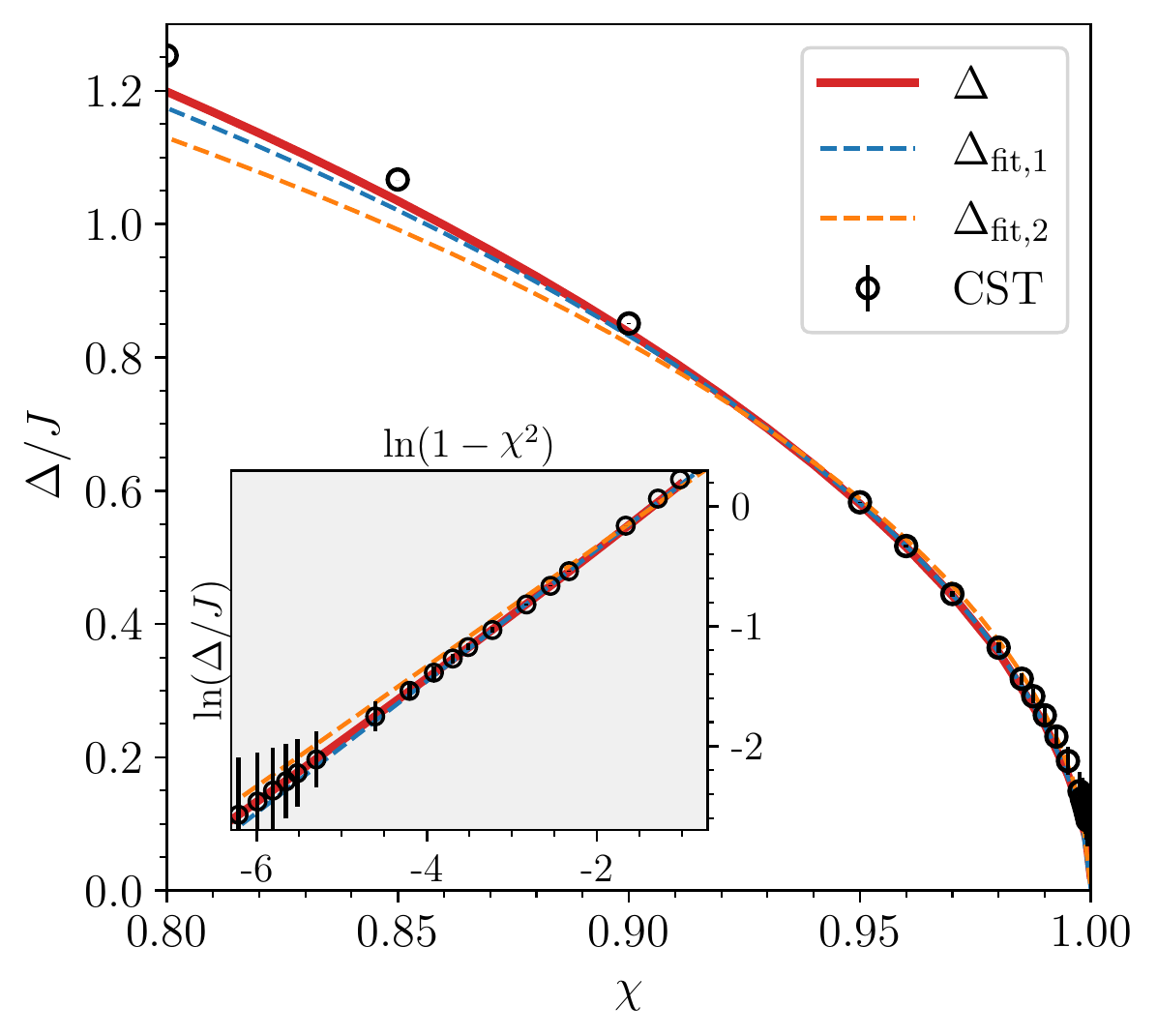}
   \caption{The spin gap $\Delta$ plotted as a function of the anisotropy $\chi$. 
	  The red solid curve shows the Schwinger boson mean-field data while the dashed curves are fits:  
		$\Delta_\text{fit,i} = c_i(1-\chi^2)^\mu_i$ in blue with $c_1=(2.04\pm 0.06)J$, $\mu_1=0.54\pm 0.02$
		and the orange curve with $\mu_2=1/2$ and $c_2=(1.88\pm 0.06)J$. The black symbols represent gap values obtained
		by CST \cite{walth23} rescaled by the factor $1.3$. 
	}
    \label{fig:AnisoGap}
\end{figure}

The behavior of the spin gap $\Delta$ as 
function of the anisotropy parameter $\chi$ is of particular interest.
The required input parameter is the product $AB$. The behavior of $A$, $B$ and of their product
 is shown in Fig.\ \ref{fig:AB} in Appendix \ref{a:magnetization-anisotropic}.
The expectation value $A$ appears to be smooth while $B$ and thus the product 
$AB$ display a singular behavior at $\chi=1$. Figure \ref{fig:AnisoGap} 
 depicts the spin gap and compares it with recently published 
data obtained by a different advanced semi-analytical approach using 
continuous similarity transformations (CST) \cite{walth23}.
The latter can be considered to be reliable. Good agreement is obtained if
the CST data is rescaled by the factor $1.3$. In view of the simplicity of the mean-field
approach, this agreement is satisfactory. In particular, mean-field and CST results are
consistent with a square root behavior $\Delta\propto (1-\chi^2)^\mu$ even though fitting
indicates slightly larger exponents $\mu\approx 0.54$.

\begin{figure}[htb]
    \centering
    \includegraphics[width=\columnwidth]{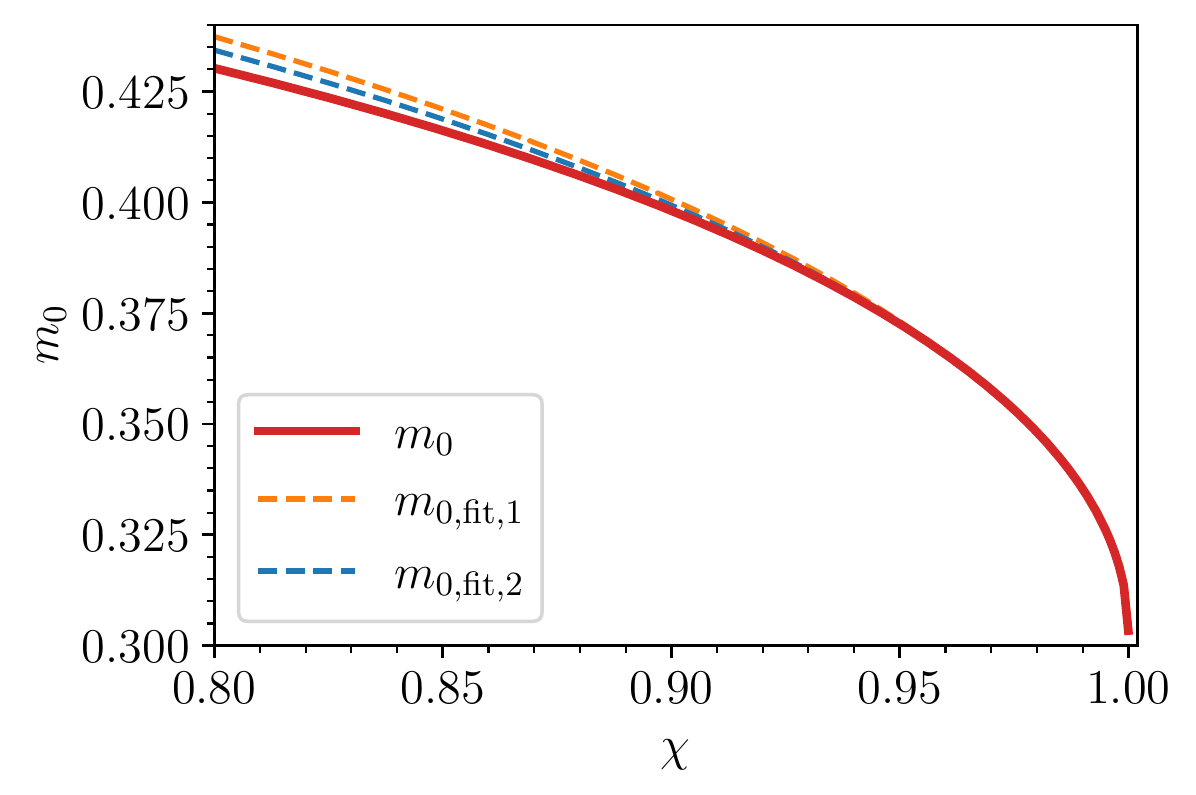}
   \caption{The magnetization at zero temperature is shown as function of the anisotropy parameter $\chi$
	according to Eq.\ \eqref{eq:magnet-aniso} as solid red line. The fit functions are
	$m_{0,\text{fit},i} = m_{0,\text{iso}} + c_i(1-\chi^2)^\mu_i$ in blue with $c_1=(0.215\pm 0.005)J$, $\mu_1=0.49\pm 0.01$
		and the orange curve with $\mu_2=1/2$ and $c_2=(0.223\pm 0.004)J$.
	}
    \label{fig:magnetization}
\end{figure}

In order to determine the magnetization we rewrite Eq.\ \eqref{eq:magnetization}
in terms of the $\alpha$- and $\beta$-particles and explicitly obtain
\be
\label{eq:magnet-aniso-sum}
m_0 = \frac{\lambda}{2N}\sumk\Bigg[\frac{\coth(\frac{\beta\omega^{-}_\mathbf{k}}{2})}{\omega^{-}_\mathbf{k}}
		-\frac{\coth(\frac{\beta\omega^{+}_\mathbf{k}}{2})}{\omega^{+}_\mathbf{k}}\Bigg] .
\ee
Its evaluation at $T=0$ is given in Appendix \ref{a:magnetization-anisotropic}
yielding
\be
\label{eq:magnet-aniso}
m_0 = S+\frac{1}{2}- \frac{1}{8\pi^2}\int_{\text{BZ}}\!\!\!dk^2 \frac{C_-}{\sqrt{C_-^2 - C_+^2 \gam^2}} .
\ee
 The explicit data is displayed
in Fig.\ \ref{fig:magnetization} as function of $\chi$ for $S=1/2$. As expected the magnetization approaches its maximum 
value $S$ for $\chi\to0$ where the model becomes the Ising model.

   \subsection{Finite temperature}
			\label{ss:finite-T}	
		
Here we turn to the effect of finite temperature. We are not interested in the phase without 
order occurring at high temperatures above the N\'eel temperature $T_\text{N}$. In the ordered phase,
 a difference between the dispersion of the $\alpha$-boson and the $\beta$-boson persists.
In contrast, however, to zero temperature no condensation of either boson occurs. 
Both bosons become gapped at finite temperature. Note that this does not imply that there is no longer
a finite sublattice magnetization $m_0$ because the anisotropy ensures that $m_0\ne 0$
is possible also up to some finite N\'eel temperature.

\begin{figure}[htb]
    \centering
    \includegraphics[width=\columnwidth]{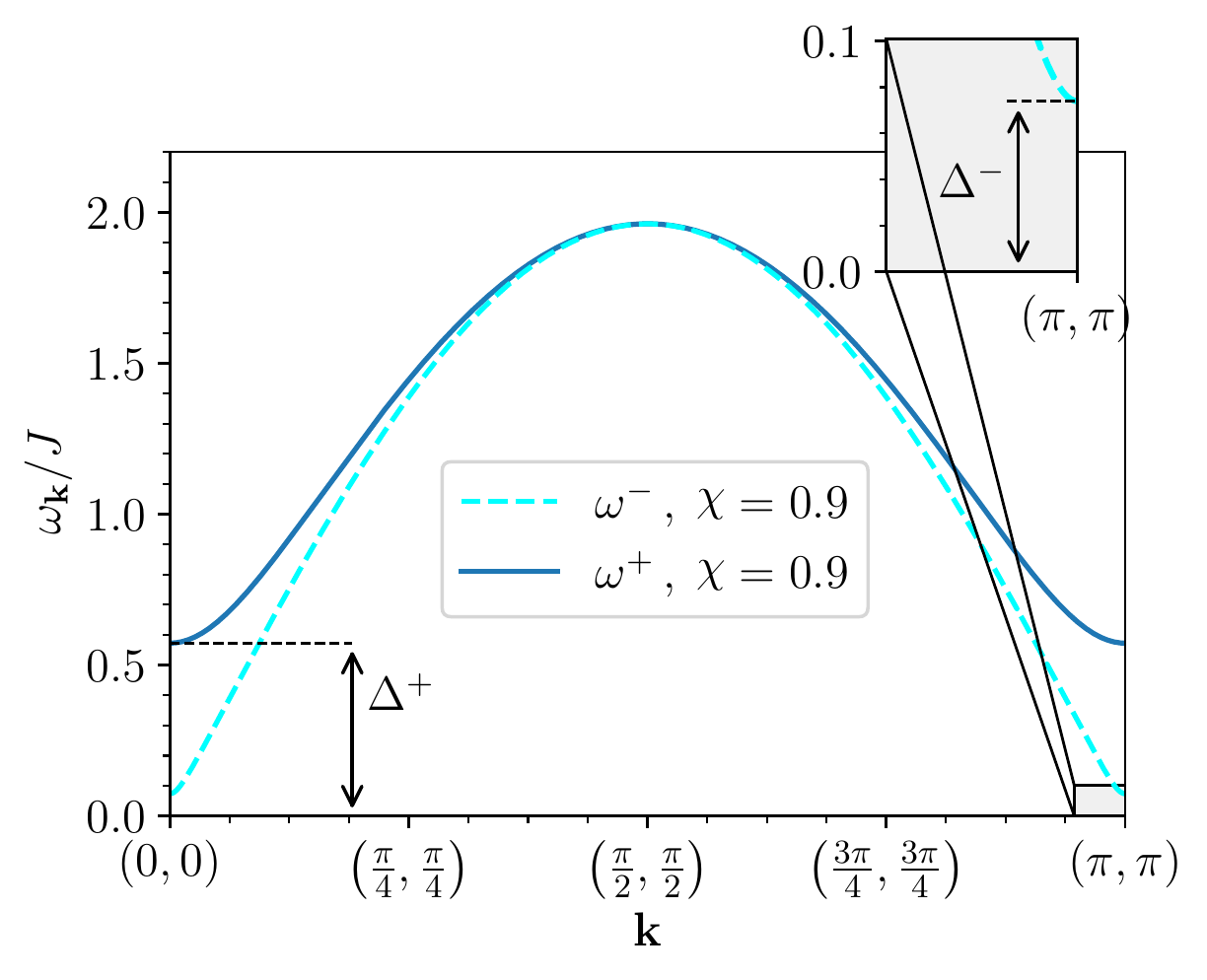}
    \caption{Dispersions $\omega^{\pm}_\mathbf{k}$ in an antiferromagnetic spin-$\frac{1}{2}$ square lattice 
		at finite temperature, plotted exemplarily for $\chi = 0.9$ at $T=0.65J$. 
    }
    \label{fig:dispersion-Tfinite}
\end{figure}

For simplicity and for later use we consider a finite mesh of the Brillouin zone on which we
evaluate the expectation values required to determine $A$, $B$, and the Lagrange parameter
$\lambda$ self-consistently, see also Sect.\ \ref{s:switch}. 
This is done for systems with linear extension up to $L=200$ at finite temperature and 
up to $L=500$ at zero temperature so that 
finite size effects are very small and negligible. In order to find the self-consistent solution 
of the ordered phase in practical implementations one must
start from initial guesses for the parameters which allow for a difference between the 
$\alpha$- and the $\beta$-boson. This means one has to start with some finite $B>0$.

\begin{figure}[htb]
    \centering
    \includegraphics[width=\columnwidth]{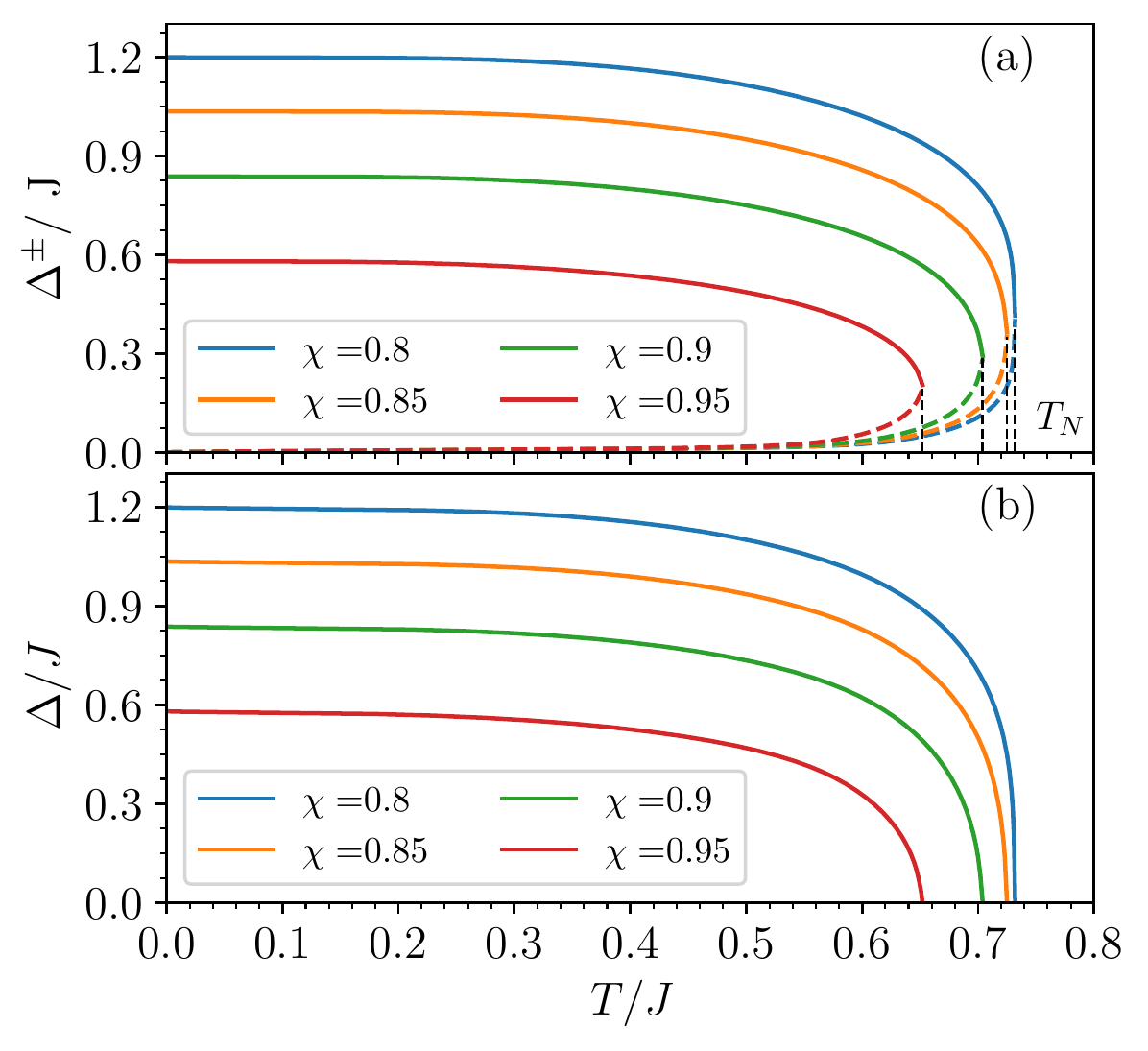}
    \caption{Panel (a): Auxiliary gaps $\Delta^\pm$ as function of temperature for various anisotropies $\chi$.
		The gap $\Delta^+$ shown by solid lines refers to the $\beta$-bosons; the gap $\Delta^-$ shown by dashed lines refers 
		to the $\alpha$-bosons.
		The two gaps merge at the N\'eel temperature $T_\text{N}$ indicated by short vertical black lines.
		Panel (b): Effective physical gap $\Delta$ according to Eq.\ \eqref{eq:phys-spin-gap}.
		}
    \label{fig:gaps-Tfinite}
\end{figure}

Fig.\ \ref{fig:dispersion-Tfinite}
displays the resulting dispersions $\omega^-_\kk$ of the $\alpha$-boson and $\omega^+_\kk$ of the $\beta$-boson.
As stated before, both are gapped. But for positive $m_0$ the dispersion $\omega^-_\kk$
is still lower than the dispersion $\omega^+_\kk$ so that more $a$-bosons are present than
$b$-bosons. We define the gaps
\bes
\begin{align}
\Delta^\pm &\coloneqq \omega^\pm_{\kk=\mathbf{0}},
\\
\Delta &\coloneqq \Delta^+  - \Delta^- .
\label{eq:phys-spin-gap}
\end{align}
\ees
The gaps $\Delta^\pm$ are auxiliary quantities without direct physical meaning because the
physical spin excitations always imply an action on two bosons. In particular, the annihilation
of an $\alpha$-boson and the creation of a $\beta$-boson represents a spin flip down for $S=1/2$
or generally a lowering of the sublattice magnetization by the creation of one magnon.
Thus, the above quantity $\Delta$ represents the physical spin gap. Figure \ref{fig:gaps-Tfinite}
displays the auxiliary gaps $\Delta^\pm$ in panel (a) and the physical gap $\Delta$
in panel (b). In a rigorous treatment, the spin gap is a property at $T=0$ referring to
the minimum energy between the ground state and the first excited state(s). In this view,
no ``temperature dependent'' spin gap makes sense. The spin gap in a mean-field theory
must be seen as an \emph{effective} spin gap which reflects the gap of the bilinear mean-field 
Hamiltonian which describes the physics of the underlying interacting model best at a given
temperature. 

\begin{figure}[htb]
    \centering
    \includegraphics[width=\columnwidth]{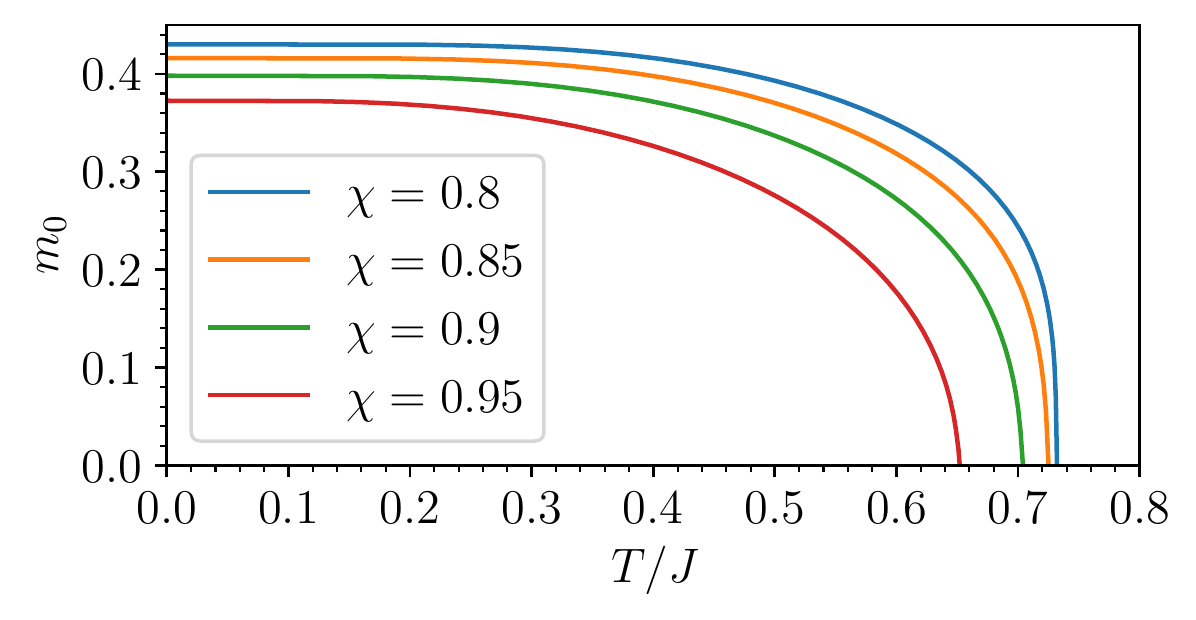}
    \caption{Sublattice magnetization $m_0$ as defined in Eq.\ \eqref{eq:magnetization} for various 
		anisotropies $\chi$ as function of temperature. The solid line is a guide to the eye.
		It vanishes at the N\'eel temperature, where the ordered phase ceases to exist, with a square root
		behavior.
		}
    \label{fig:magnet-Tfinite}
\end{figure}

Once we know all the expectation values in reciprocal space we also know all expectation values
in real space. Thus the sublattice magnetization can also be computed and it is displayed
as function of temperature in Fig.\ \ref{fig:magnet-Tfinite} up to the respective
N\'eel temperatures. The power law close to the N\'eel temperatures is expected to be a square root
law as is generic for mean-field theories. Our data is fully consistent with this assumption. 
The analogous square root law is consistent with our findings for the effective spin gap $\Delta$
in Fig.\ \ref{fig:gaps-Tfinite}(b).

\begin{figure}[htb]
    \centering
    \includegraphics[width=\columnwidth]{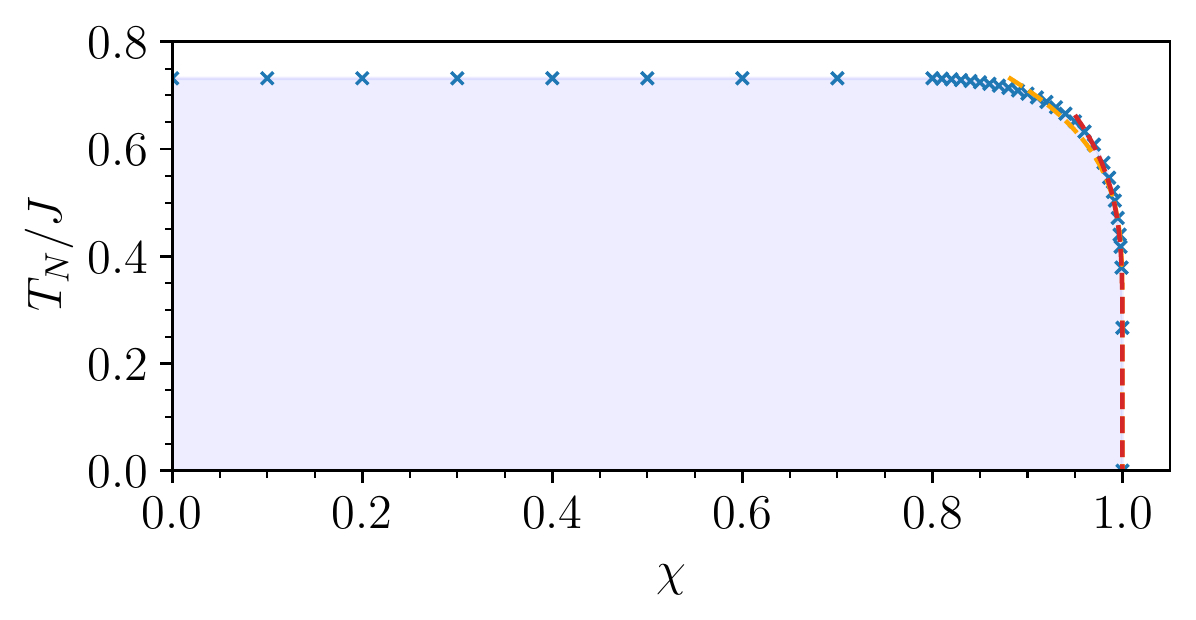}
    \caption{N\'eel temperature as function of the anisotropy. In accord with the Mermin-Wagner theorem,
		the N\'eel temperature vanishes in the isotropic case, i.e., for $\chi\to 1$.
		The vanishing of $T_\text{N}$ close to the isotropic point can be fitted according to
		$T_\text{N}/J \approx c_1/(|\ln(1-\chi)|+c_2)$ with $c_1=4.1\pm 0.08$ and $c_2=3.5\pm 0.02$ in the
		interval $\chi\in[0.88,1]$ (orange dashed line) and with $c_1=3.5\pm 0.05$ and $c_2=2.3\pm 0.06$ in the
		interval $\chi\in[0.95,1]$ (red dashed line). 
				}
    \label{fig:neel-temp}
\end{figure}

Finally, Fig.\ \ref{fig:neel-temp} depicts the N\'eel temperature as function of
the anisotropy according to the Schwinger boson mean-field theory. As pointed out before
this mean-field theory complies with the rigorous Mermin-Wagner theorem \cite{mermi66} so that
$\lim_{\chi\to1} T_\text{N}(\chi) = 0 $ holds. But the downturn upon approaching
the isotropic case is extremely steep. From the integrals in the derivation of the Mermin-Wagner
theorem for two dimensions a logarithmic dependence according to $T_\text{N}/J \approx c_1/(|\ln(1-\chi)|+c_2)$
appears plausible, see caption. 
To estimate the accuracy of our results we can compare them at $\chi=0$ with the rigorous
result of Onsager \cite{onsag44} for the two-dimensional Ising model yielding $T_\text{N}=0.5673J$. 
As for the gap, the results agree if the mean-field results are scaled down by a factor of $1.3$.

The results presented in this section provide an overview of the essential properties
of the system in equilibrium. It is crucial to know them quantitatively because
they define the initial conditions for the subsequent time-dependent solutions
which describe the switching process.

    \section{Switching the sublattice magnetization}
    \label{s:switch}
		
		The aim is to invert the sublattice magnetization $m_0\to -m_0$. We denote the time-dependent
		sublattice magnetization during the switching process by $m(t)$. One can 
		think of this process as switching a bit from its 1-state to its 0-state. Hence,
		the considered process is highly relevant in data storage with the advantages
		exposed in the Introduction. 
		
		We have a precessional rotation in mind. Thus we add a static magnetic field along
		the $S^y$-axis in form of a Zeeman term from time $t=0$ onwards
		\bes
		\label{eqn:noneqisoH}
		\begin{align}
    \mathcal{H} &= \mathcal{H}_{0} - h_y\sum_i S_i^y \\
		&= \mathcal{H}_{0} - \frac{h_y}{2i}\sum_i \lr{S_i^+-S_i^-} \\
		&= \mathcal{H}_{0} - \frac{h_y}{2i}\sum_i (a_i^\dagger b_i - b_i^\dagger a_i),
		\end{align}
\ees
where $\mathcal{H}_{0}$ is the unperturbed Hamiltonian \eqref{eqn:SchwingerH3}.

For the isotropic case, the Zeeman term does not change under the
Bogoliubov transformation yielding
\begin{align}
\nonumber
    \mathcal{H} &= E_\text{MF} - N\lambda + \sumk \omega^\text{iso}_\mathbf{k}\lr{\nbk{\alpha} + \nbk{\beta} +1} -
		\\ &\qquad 
		\frac{h_y}{2i}\sumk\lr{\alk^\dagger\bek - \bek^\dagger\alk}.
\end{align}
Due to the spin isotropy the Zeeman term commutes with the Hamiltonian so that any ground state
of $\mathcal{H}_{0}$ remains a ground state under the action of the Zeeman term. 
The Zeeman term induces a collective rotation about $S^y$ at all sites
simultaneously.
The rotation of each spin can be treated as if the spin were isolated.
Hence, the sublattice magnetization is rotated in the $S^zS^x$-plane. 
A rotation about the angle $\varphi$ is achieved within the time interval 
\be
\label{eq:switch-time}
t_\varphi=\frac{\varphi}{h_y}.
\ee
This can be verified in terms of the spin operators or in terms of the Schwinger bosons.
The tilt $m_0\to-m_0$ is achieved for $t_\pi={\pi/h_y}$. No minimum magnetic field
is required to achieve the rotation if one can create long enough pulses of constant magnetic 
fields. Yet, this case is not promising for application because the absence of any
anisotropy also implies that the system does not have any rigidity of the sublattice magnetization against
perturbations. 

For this reason, we turn to the anisotropic case where the situation is
more subtle because the Zeeman term does not commute with the Hamiltonian.
We have to compute the time evolution under the full mean-field Hamiltonian
\begin{align}
\nonumber
   \mathcal{H} &= E_\text{MF} -\frac{1}{2}\sumk \gam \Big(C_-\ak^\dagger \akm^\dagger + C_+\bk^\dagger\bkm^\dagger
	\\ \nonumber
	& \qquad + C_-\ak\akm + C_+\bk\bkm\Big) + \lambda \sumk \lr{\nbk{a} + \nbk{b}} 
	\\
	& \qquad - \frac{h_y}{2i}\sumk\lr{\ak^\dagger\bk - \bk^\dagger\ak}.
	\label{eq:hamilton-switch}
\end{align} 
We stress that the prefactors $C_\pm$ defined
in Eq.\ \eqref{eq:C-def} depend on expectation values which themselves acquire a temporal
dependence upon switching. Hence the above Hamiltonian itself is time dependent. 
We refrain from expressing the Hamiltonian \eqref{eq:hamilton-switch} in terms of
Bogoliubov particles $\alk$ and $\bek$ because no useful simplification can be reached due to the 
time dependence of the $C_\pm$. It would require to adjust the Bogoliubov transformation
at each instant of time in order to keep diagonality. This is tedious and does not
promise neither insight nor advantage.

The Hamiltonian \eqref{eq:hamilton-switch} is fully sufficient to compute the
time dependence of the expectation values by means of Heisenberg's equations of
motion. This amounts up to the density-matrix formalism. The set of differential equations
reads
\begin{subequations}
\begin{align}
\label{eqn:DissEQ1}
\begin{split}
    \partial_t \lara{\ak^\dagger \ak} &= 2 \gam \Im\big(C_-^*\lara{\ak\akm}\big) \\\
		 & \qquad + \frac{h_y}{2}\big(\lara{\ak^\dagger\bk}+\lara{\bk^\dagger\ak}\big), 
		\end{split}
\\
\begin{split}
    \partial_t \lara{\bk^\dagger \bk} &= 2 \gam \Im\big(C_+^*\lara{\bk\bkm}\big) \\\ 
    & \qquad - \frac{h_y}{2}\big(\lara{\ak^\dagger\bk}+\lara{\bk^\dagger\ak}\big),\label{eqn:DissEQ2}
\end{split}
\\
\begin{split}
    \partial_t \lara{\ak^\dagger \bk} &= - i \gam \big(C_-^*\lara{\ak\bkm} - C_+\lara{\ak^\dagger\bkm^\dagger}\big)\\\
		&\qquad + \frac{h_y}{2}\big(\lara{\bk^\dagger\bk}-\lara{\ak^\dagger\ak}\big) ,\label{eqn:DissEQ3}
\end{split}
\\
\begin{split}
    \partial_t \lara{\ak \akm} &=  i \gam \big[C_- (2\lara{\nbk{a}}+1)\big]\\\ 
		&\qquad  - 2 \lambda  i \lara{\ak\akm} + h_y\lara{\ak\bkm} ,\label{eqn:DissEQ4}
\end{split}
\\
\begin{split}
    \partial_t \lara{\bk \bkm} &=  i \gam \big[C_+ (2\lara{\nbk{b}}+1)\big]\\\
		&\qquad - 2 \lambda  i \lara{\bk\bkm} - h_y\lara{\ak\bkm} ,\label{eqn:DissEQ5}
\end{split}
\\
\nonumber
    \partial_t \lara{\ak \bkm} &=  i \gam \big( C_-\lara{\ak^\dagger\bk} + C_+ \lara{\ak^\dagger\bk}^* \big) 
		-2 \lambda  i \lara{\ak\bkm}\\
    & \qquad  - \frac{h_y}{2}\big(\lara{\ak\akm}-\lara{\bk\bkm}\big).
				\label{eqn:DissEQ6}
\end{align}
\end{subequations}

\begin{figure}[htb]
    \centering
     \includegraphics[width=\columnwidth]{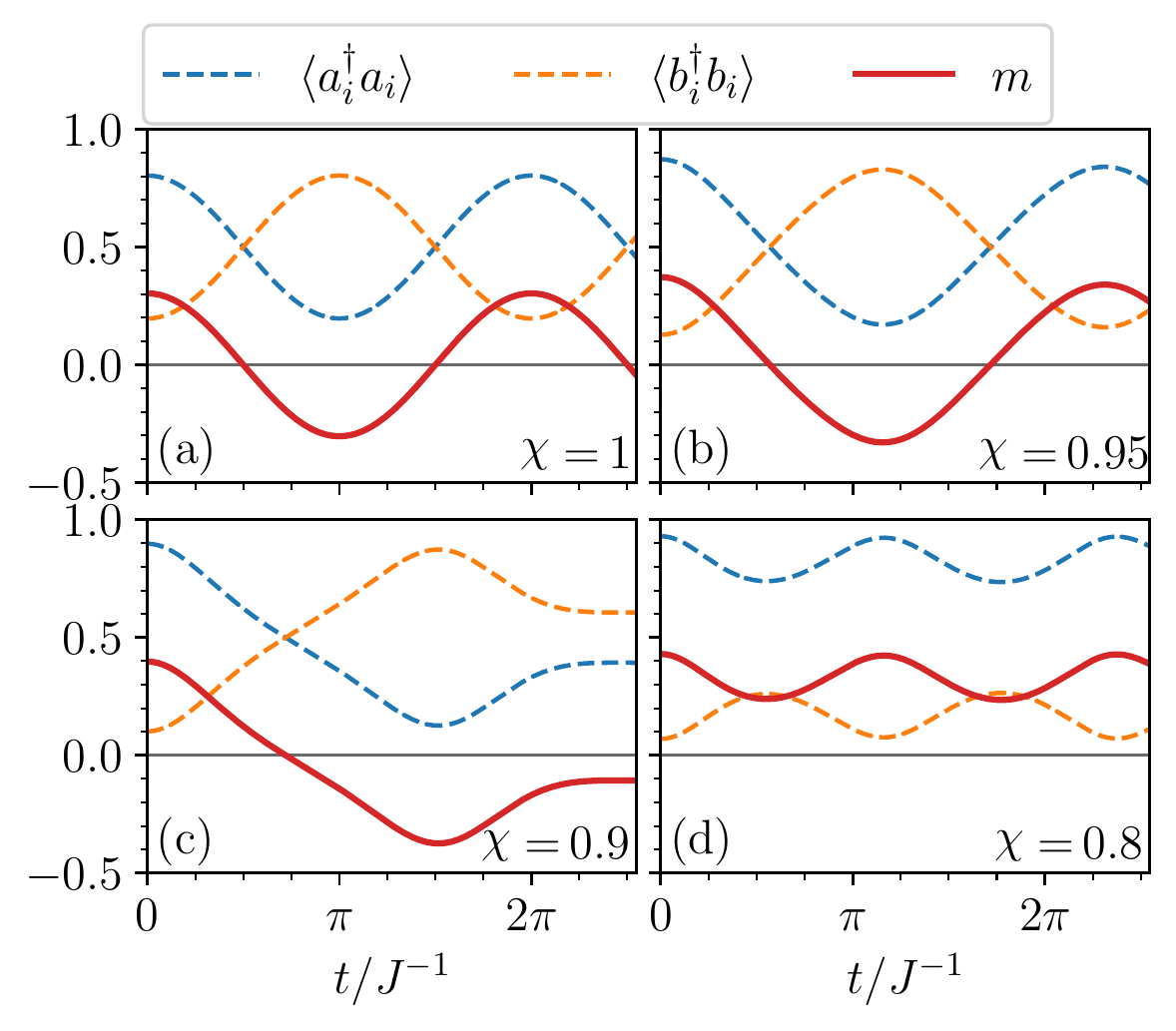}
    \centering
		\caption{Dynamics of the occupation of the Schwinger bosons $\lara{\nbi{a}}$
    and $\lara{\nbi{b}}$, as well as the resulting magnetization $m$ in the time interval 
		$t \in [0\, J^{-1}, 8\, J^{-1}]$ for exemplary values of the anisotropy.
    The strength of the applied field is $h=1\,J$ so that in the isotropic case (panel (a), $\chi=1$), the switching 
		duration of $t = \pi  J^{-1}$ for the $\pi$-pulse results in accord with Eq.\ \eqref{eq:switch-time}. 
    The calculations were performed for a system size of $L=500$ implying $N=250,\!000$.
		}
    \label{fig:switch1}
\end{figure}

In terms of the initial Schwinger bosons without any Bogoliubov transformations, the
relations defining $A$, $B$, and the constraint read
\bes
\begin{align}
 A &= \lara{a_ia_j} + \lara{b_ib_j} = \frac{1}{N}\sumk \gam\lr{\lara{\ak\akm}+\lara{\bk\bkm}} , \label{eqn:IV3} \\ 
 B &= \lara{a_ia_j} - \lara{b_ib_j} = \frac{1}{N}\sumk \gam\lr{\lara{\ak\akm}-\lara{\bk\bkm}} , \label{eqn:IV4}\\
2S &= \lara{\nbi{a}} + \lara{\nbi{b}} = \frac{1}{N}\sumk\big(\lara{\nbk{a}} + \lara{\nbk{b}}\big) , \label{eqn:IV5}
\end{align}
\ees
where $i,j$ are adjacent sites. This completes a closed set of differential equations. The constraint
\eqref{eqn:IV5} is not needed for the temporal equation. But we checked that it is always fulfilled
in the course of time if it is fulfilled initially as is ensured by starting from a valid equilibrium solution. 

\subsection{Switching at zero temperature}
\label{ss:switch-Tzero}

Figure \ref{fig:switch1} displays the dynamics of the occupations
of the Schwinger bosons and the resulting sublattice magnetization $m(t)$ given by
\be
m(t) = \big(\lara{\nbi{a}} - \lara{\nbi{b}}\big)/2 = \frac{1}{2N}\sumk\big(\lara{\nbk{a}} - \lara{\nbk{b}}\big) 
\ee
for various values of the anisotropy parameter $\chi$. Panel (a) shows the result for the isotropic case. The
external field applied perpendicular to the sublattice magnetization $m_0$ nicely rotates it following a cosine-curve
to $-m_0$ at time $t_\pi$ predicted by Eq.\ \eqref{eq:switch-time}. Since the switching field persists
the magnetization is switched back to the initial value at $t=2t_\pi$. No reduction of $m$ due to dephasing
is discernible; the rotation is fully coherent.

Panel (b) reveals differences to the isotropic case. First, the initial magnetization is larger in 
agreement with the results shown in Fig.\ \ref{fig:magnetization}. The switching to negative 
values of $m$ succeeds, but it takes longer than in the isotropic case. We attribute this
to the anisotropy which hinders the rotation to take place although the state with sublattice
magnetization $-m_0$ is also a valid ground state. But the states in between are neither ground states
nor eigenstates of the system. This leads to dephasing of the modes at different wave vectors and 
implies that $|m(t)|$ does not reach the initial value $m_0$ anymore.

Increasing the anisotropy by lowering $\chi$ we see an even longer
switching process in panel (c) of Fig.\ \ref{fig:switch1}. Surprisingly, only one switching appears
to be possible since the switching back to the original sign of $m$ does not take place.
The switched modes appear to be out of phase, i.e., dephasing has been quite detrimental.
We emphasize that such effects are not captured by the description of the sublattice magnetization
by a classical vector.
Finally, panel (d) displays an example where no switching
occurs at all. The sublattice magnetization oscillates only a little below its initial
value $m_0$. These observations show that a minimum magnetic field
is required in order to change the sign of $m$. The anisotropy generates a degree of robustness
which needs to be overcome by the external magnetic field.

\begin{figure}[htb]
        \centering
       \includegraphics[width=\columnwidth]{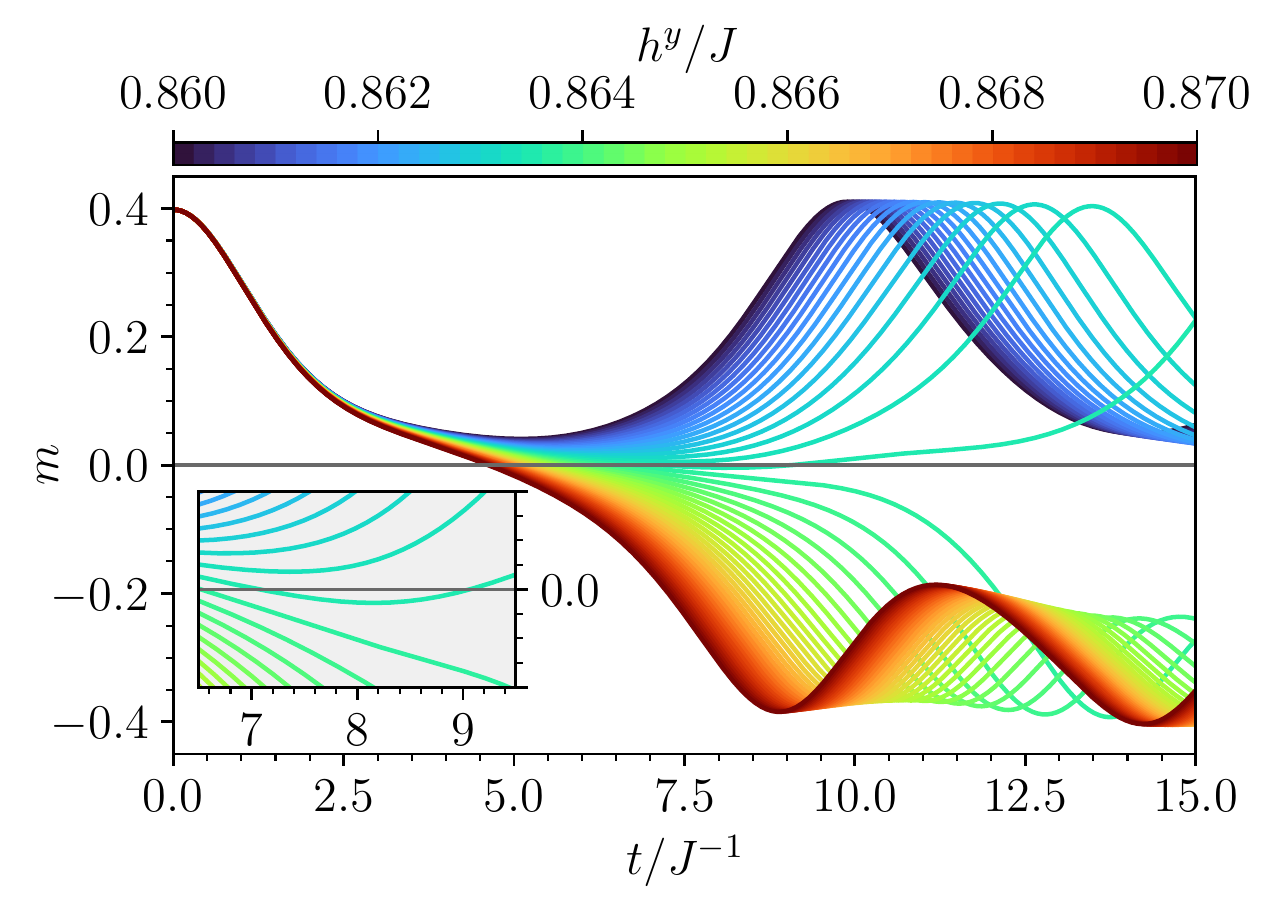}
       \caption{Time evolution of the magnetization $m$ for various external fields 
			  $h^y\in[0.86\,J,\,0.87\,J]$ at the anisotropy $\chi=0.9$.
        The color bar on top indicates the external 
        field. Clearly, qualitatively distinct temporal behaviors occur
				depending on the strength of the applied magnetic field.
				There is a threshold value $h_\text{t}$
				above which switching is possible and below which no switching is possible.
				}
        \label{fig:scan-magnet-field}
    \end{figure}				
				
In order to further investigate the conditions for successful switching  we scan $m(t)$ for
a range of applied magnetic fields in Fig.\ \ref{fig:scan-magnet-field}. We observe a distinct
difference in the behavior for lower fields compared to the behavior for larger ones. Large ones
enable switching and low ones do not. There is a well-defined value $h_\text{t}$ 
separating the two regimes. We define a criterion to distinguish whether switching is possible or not.
The occurrence of a negative value of $m(t)$ is not an ideal criterion as the inset
of Fig.\ \ref{fig:scan-magnet-field} shows. Instead, we choose the occurrence
of an inflection point before the first extremal value for $t>0$ as criterion. If such
an inflection point exists $m(t)$ continues to turn down reaching
a substantial negative value. Otherwise, no switching is possible.

\begin{figure}[htb]				
        \centering
        \includegraphics[width=\columnwidth]{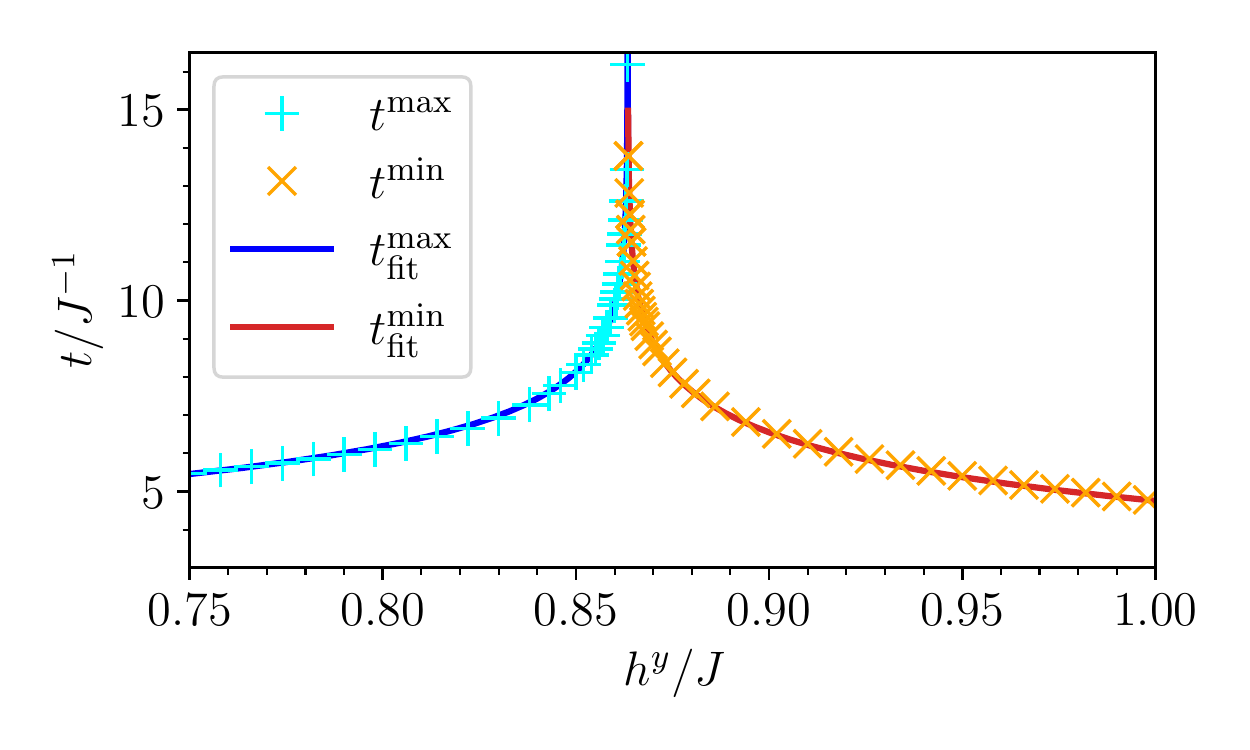}
        \caption{Times (symbols) at which $m(t)$ reaches the relevant
        minimum or maximum, depending on whether switching is possible (orange) or not (cyan)
				plotted against the external field $h^y$. The position of the singularity defines
				the threshold field $h_\text{t}$.
        More data points were computed around the singularity to obtain a higher 
        accuracy for fitting the function given in Eq.\ \eqref{eq:fit}.
        }
        \label{fig:Hthres2}
    \end{figure}

Next, we quantify how long the switching takes. The duration $t$ is
the instant in time when the negative minimum is reached. 
If no switching is possible, we define the instant in time when the positive
maximum is reached as the duration of the failed switching attempt.
These durations are plotted in Fig.\ \eqref{fig:Hthres2} together with
suitable logarithmic fits
\be
\label{eq:fit}
t_\text{fit}(h^y) = c_1 + c_2 \ln|(h^y - h_\text{t})/J|.
\ee
The fits describe the data remarkably well below and above the threshold with very similar
parameters for the minima (succeeded switching)
\bes
\label{eq:succeeded}
\begin{align}
   c_1^\text{min} &= (2.04 \pm 0.06)J^{-1}, \\
    c_2^\text{min} &= (-1.36   \pm 0.01) J^{-1},\\
    h_\text{t}^\text{min} &= (0.8634 \pm 5 \cdot 10^{-6} ) J,
\end{align}
\ees
and for the maxima (failed switching)
\bes
\label{eq:failed}
\begin{align}
c_1^\text{max} &= (2.73 \pm 0.08) J^{-1},  \\
 c_2^\text{max} &= (-1.25 \pm 0.01) J^{-1}, \\
h_\text{t}^\text{max} &= (0.8634 \pm 6 \cdot 10^{-7}) J. 
\end{align}
\ees
Clearly, the resulting threshold value for $\chi=0.9$ is $(0.8634 \pm 5 \cdot 10^{-6} ) J$
which is quite substantial and fairly close to the value of the spin gap $\Delta$.
The question suggests itself why such a logarithmic divergence occurs.
We argue that this kind of divergence is a clear signature that the magnetic order
in antiferromagnets has a certain inertia in its dynamics. This was observed in experiment and
supported by a calculation based on classical equations of motion for the
antiferromagnetic vector describing the size and direction of the sublattice magnetization
\cite{kimel09}. The fact that the microscopic spin-wave description reproduces
this behavior corroborates the idea of an inertia of the antiferromagnetic order
convincingly. In \mbox{Appendix \ref{a:barrier}} we show that a  classical
motion of a massive particle over an activation barrier reproduces the logarithmic
singularity which we observed. This supports the above interpretation.

\begin{figure}[htb]
    \centering
    \includegraphics[width=\columnwidth]{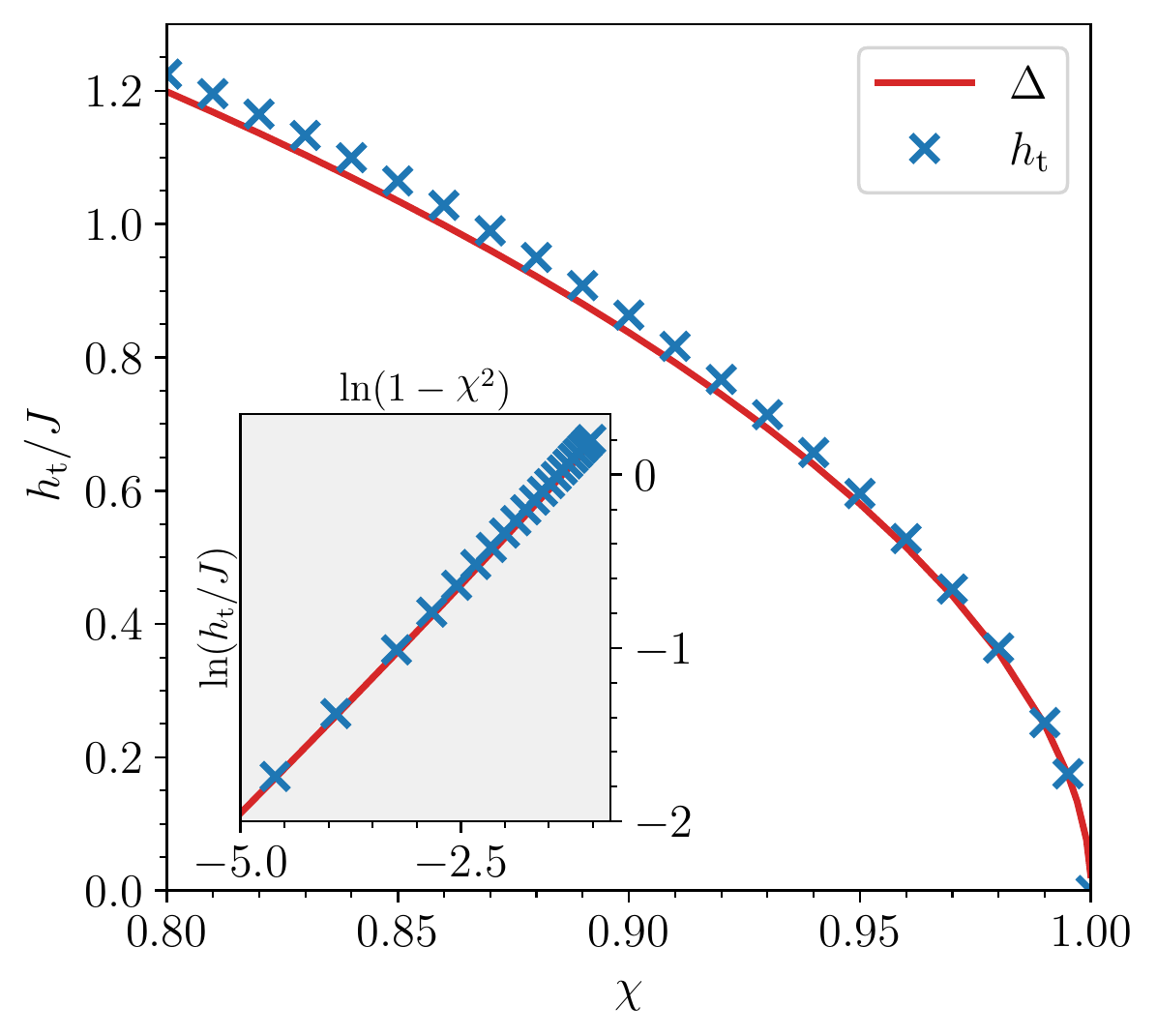}
    \caption{Threshold magnetic field $h_\text{t}$ depicted by the blue symbols vs.\ anisotropy $\chi$.
		The data is obtained for a system size $L=500$.
     The red solid curve reproduces the spin gap from Fig.\ \ref{fig:AnisoGap} to illustrate that
		the threshold magnetic fields are essentially determined by the size of the spin gap.
		}
    \label{fig:Schwellfeld}
\end{figure}

We determine these threshold values of a range of anisotropies by bisection with high
accuracy. The resulting data is shown in Fig.\ \ref{fig:Schwellfeld} by the symbols.
We compare them with the values of the spin gap shown above in Fig.\ \ref{fig:AnisoGap}.
They are very close to each other for a range of anisotropies. Thus, we conclude that
it is the size of the spin gap which determines the typical field strength required
for switching. This is in line with the idea that the spin gap measures the robustness
of the system against any kind of perturbation.

\begin{figure}[htb]
        \centering
        \includegraphics[width=\columnwidth]{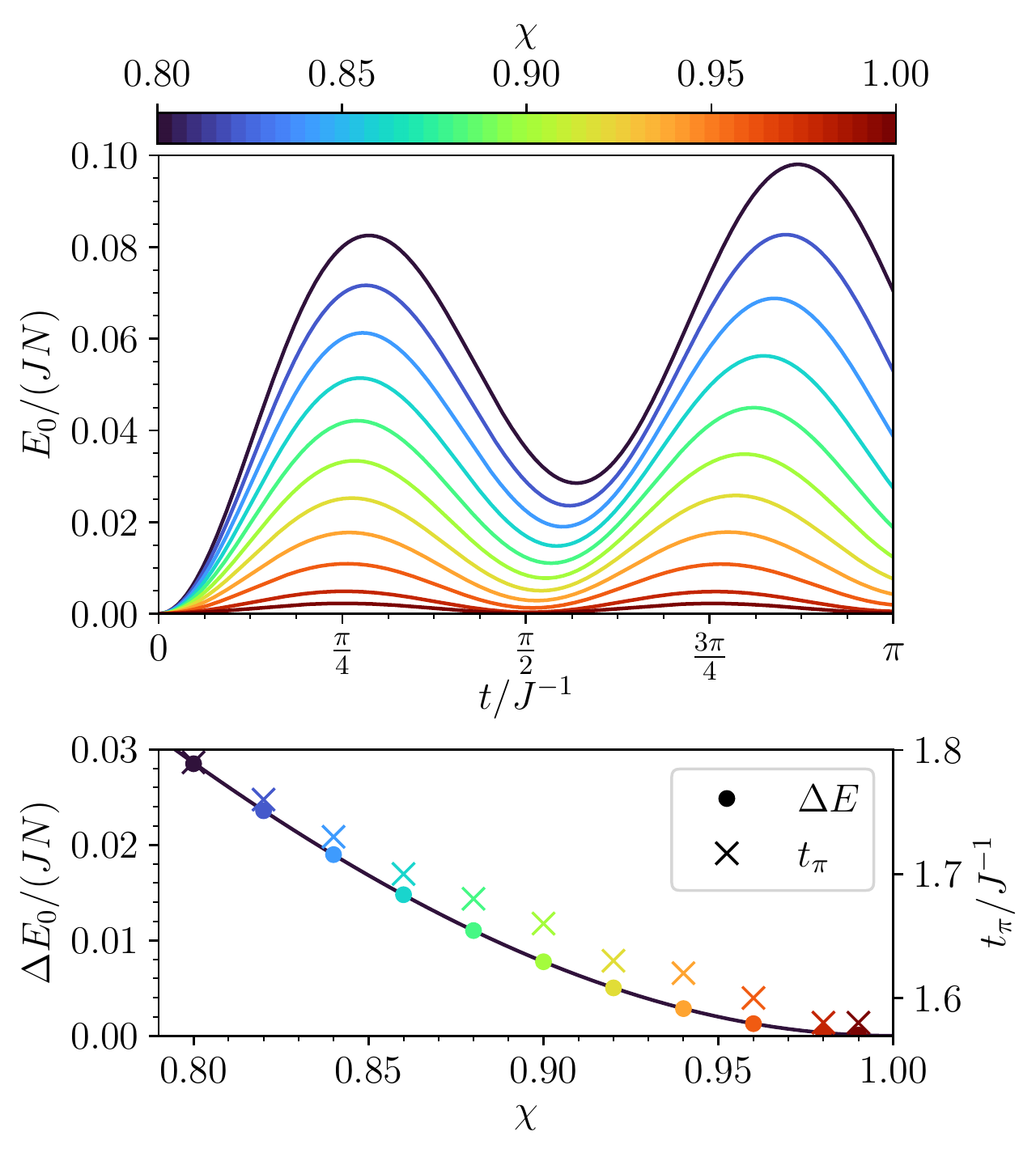}
        \caption{In the upper panel, the temporal evolution of the energy $E_0$  per 
        lattice site is shown for an external field $h=2 \,J$ and various anisotropies 
				$\chi \in [0.8,\, 0.99]$. The color bar indicates the value of anisotropy.
        The energy minima $\Delta E_0$ show the energy after switching the magnetization by $180^\circ$;
				they are not zero but indicate an energy increase because the switching does not
				lead to the degenerate ground state with magnetization $-m_0$.
				        The energy increase  after switching as well as the switching 
        duration $t_\pi$ are plotted against the anisotropy in the lower panel.
        The solid line results from a parabolic fit of the energy increase.
				}
        \label{fig:energy}
					\end{figure}

We emphasize that a key aspect of the description of the switching dynamics
in terms of spin-waves is that the contribution of each mode is captured individually.
The whole process is not one single collective motion, but it consists of the
contributions of a thermodynamically large number of modes. Hence, the coherence
is not preserved in the course of the switching process except in the isotropic 
case. This could be inferred already from the decrease of the maximum values
of $|m(t)|$ in Fig.\ \ref{fig:switch1}, i.e., $|m(t)| < m_0$ except for $t=0$.
To underline this aspect further Fig.\ \ref{fig:energy} displays the energy in the course
of the switching measured by the expectation value of $\mathcal{H}_0$
\begin{align}
    E(t) &= N\Big\{2S\lambda + 2S^2 -\lara{a_ia_j}\lara{a_i^\dagger a_j^\dagger} 
		- \lara{b_ib_j}\lara{b_i^\dagger b_j^\dagger}\nonumber \\
    &\quad  - \chi\lr{\lara{b_ib_j}\lara{a_i^\dagger a_j^\dagger}+\lara{b_i^\dagger b_j^\dagger}\lara{a_i a_j}}\Big\} .
\end{align}
The dynamics of the energy relative to the energy of the initial 
state $E_0(t) \coloneqq E(t) - E(0)$ is depicted in Fig.\ \ref{fig:energy} 
in the time interval $t \in [0, \pi/J]$ 
for various degrees of anisotropy $\chi \in [0.8,\, 0.99]$.
The magnetic field was chosen relatively high at $h=2\,J$ so that a rotation 
is possible for all anisotropies in the studied  range.

As expected, there is an energy minimum after switching 
since at this instant of time the switched state is closest to the other ground state
with magnetization $-m_0$.
In the isotropic case $\chi=1$, the first minimum is reached at the time 
$t=\pi/(2 J)$ according to Eq.\ \eqref{eq:switch-time}.
However, as already seen in Fig.~\ref{fig:switch1}, the switching 
duration increases and the minimum shifts to larger times upon increasing anisotropy (lowering of
$\chi$). It is obvious that the energy of the initial state is not reached again
so that the overall energy is higher after switching the sublattice magnetization for $\chi <1$.
This underlines the effect of the many contributing spin modes which 
dephase and thereby depart from the initial coherent ground state.

The lower plot in Fig. \ref{fig:energy} shows the increase in energy $\Delta E_0$ 
after switching and the time $t_\pi$ at which the energy minimum is reached
as function of $\chi$.
The data points for $\Delta E_0$ can be approximated to high accuracy 
by a parabola $\Delta E_0(\chi) = c_0 (1-\chi^2)^m$ with the parameters
\begin{subequations}
\begin{align}
    c_0 &= (0.230 \pm 0.002)J  , \\
    m &= 2.04 \pm 0.05 .
\end{align}
\end{subequations}
The switching times also increase steadily with growing anisotropy, 
but they do not evolve completely smoothly in our simulation, 
especially for $\chi \to 1$. This is to be attributed to the 
finite discretization used in the numerical computations.

\subsection{Switching at finite temperature}
\label{ss:switch-Tfinite}

So far, we studied the conditions for switching the magnetization at zero temperature.
But the effect of finite temperature is highly relevant for two reasons. First, having
applications in mind, a finite temperature must be accounted for because no setup will
be operated at zero temperature. Second, 
temperature is a parameter which reduces the effective spin gap and makes the
ordered system less robust and thus easier to switch. For this reason, 
a study of finite temperatures is in order.

\begin{figure}[htb]
        \centering
        \includegraphics[width=\columnwidth]{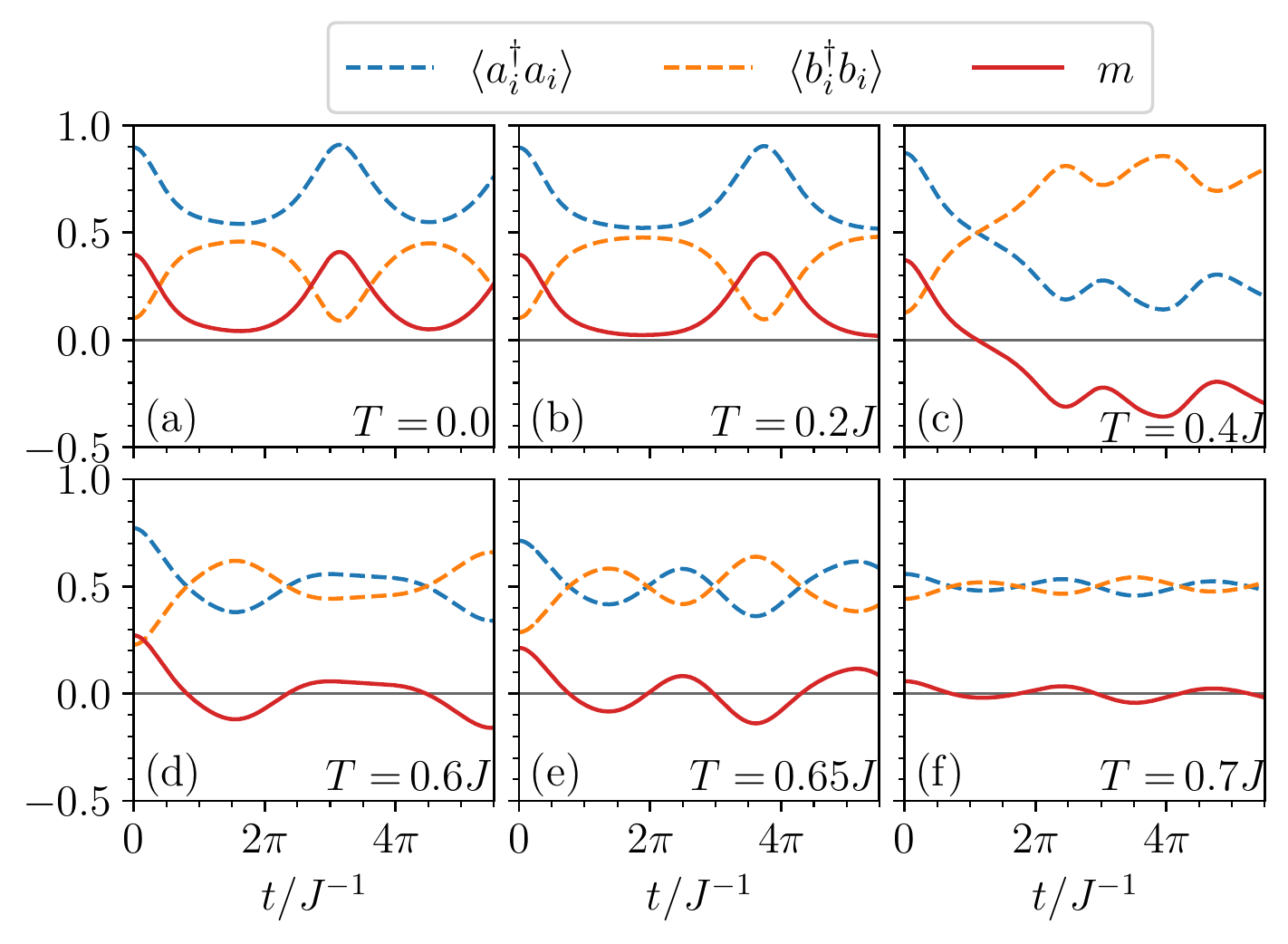}
        \caption{Same as Fig.\ \ref{fig:switch1}, but at various temperatures below the 
				N\'eel temperature $T_\text{N}=0.704J$ for $L=200$.
				}
        \label{fig:switch-Tfinite}
\end{figure}

Figure \ref{fig:switch-Tfinite} shows the effect of finite temperature
for an exemplary set of parameters.
Clearly, no switching is possible at zero and at low temperatures. But for
intermediate temperatures $T\gtrapprox 0.4J$, at least one swap $m_0\to -m_0$
is possible. Approaching the N\'eel temperature switching becomes possible
even multiple times. Of course, the switched sublattice magnetization 
is reduced in its absolute value upon approaching $T_\text{N}$. But it can still
be manipulated.

This is indeed a very promising observation because
it suggests that temporary heating of the system up to the vicinity
of the critical temperature, while staying still below it,
facilitates the writing process of information into a long-range ordered
magnetic system. For long time storage, the temperature can be lowered again
after the writing process.

\begin{figure}[htb]
        \centering
        \includegraphics[width=\columnwidth]{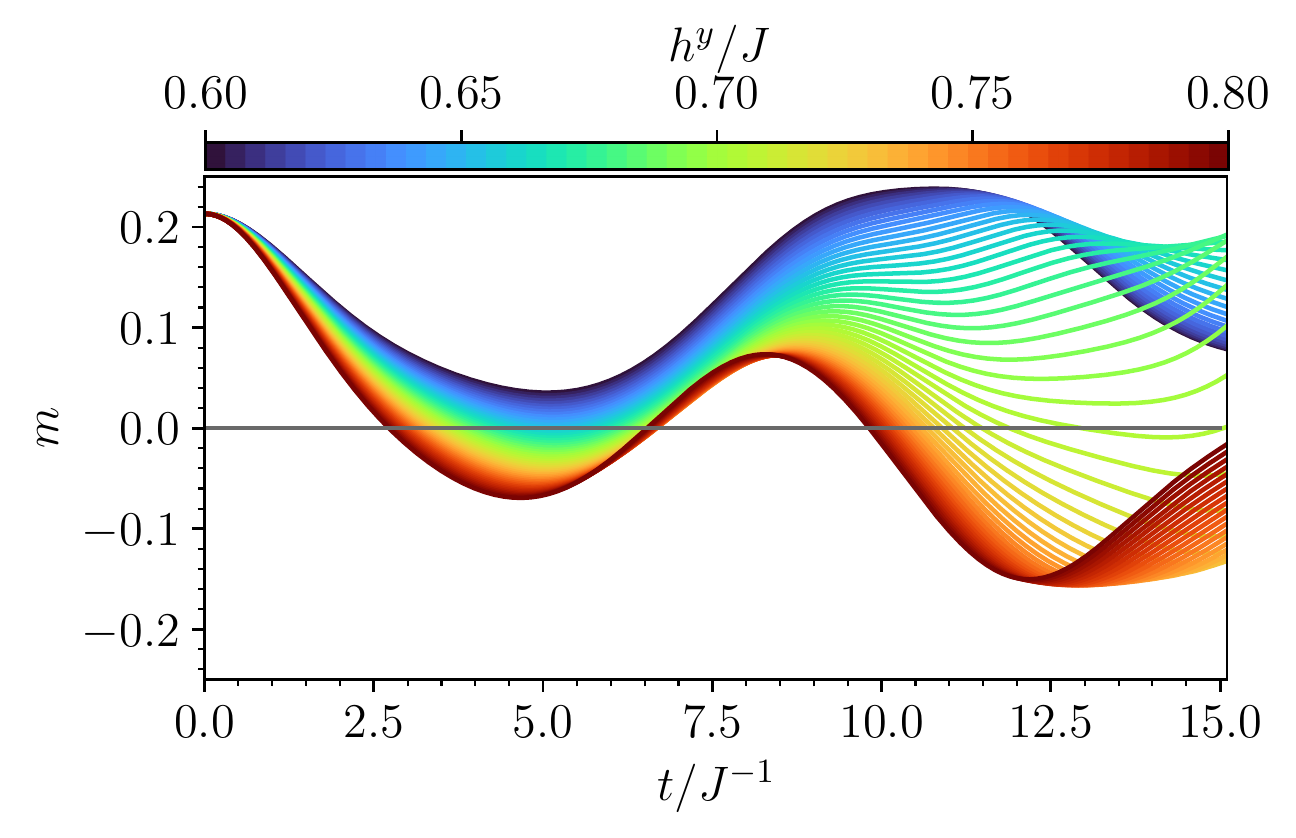}
        \caption{Time evolution of the magnetization $m$ for various external fields 
			  $h^y\in[0.6J, 0.8J]$ at anisotropy $\chi=0.9$ and temperature $T=0.65\,J$
				below $T_\text{N}$. The color bar on top indicates the external 
        field. We determine the threshold value $h_\text{t}$ from the magnetic field
				at which the first minimum of $m(t)$ touches $m=0$.}
        \label{fig:switch-Tfinite2}
\end{figure}

Hence, we investigate the finite temperature case further.
Figure \ref{fig:switch-Tfinite2} displays the temporal evolution
of $m(t)$ for various applied magnetic fields for a generic set 
of anisotropy and temperature. The
phenomenology is similar, but not identical to the one
at zero temperature, cf.\ Fig. \ref{fig:scan-magnet-field}.
At zero temperature, the switched curves appear to be
mirror images of the non-switched curves flipped around $m=0$.
At finite temperature, it is mostly the first minimum of $m(t)$ which
decreases further and further upon increasing the control field $h^y$.
Here, we take the occurrence of a negative value of $m(t)$ as signature of 
switching, i.e., the threshold field $h_\text{t}$ is determined
from the field at which the first minimum touches the $m=0$ line.
We point out that the threshold field determined in such a way
can depend on the considered time interval,
see also below, in particular if the instant in time at
which the magnetization switches sign jumps as function of the
applied field. We analyzed the time interval $t\in [0,15/h^y]$, i.e, 
for low values of the field we scanned large intervals.
Studying even larger intervals can only lower the values for $h_\text{t}$
so that our values are at least rigorous upper bounds.

\begin{figure}[htb]
        \centering
        \includegraphics[width=\columnwidth]{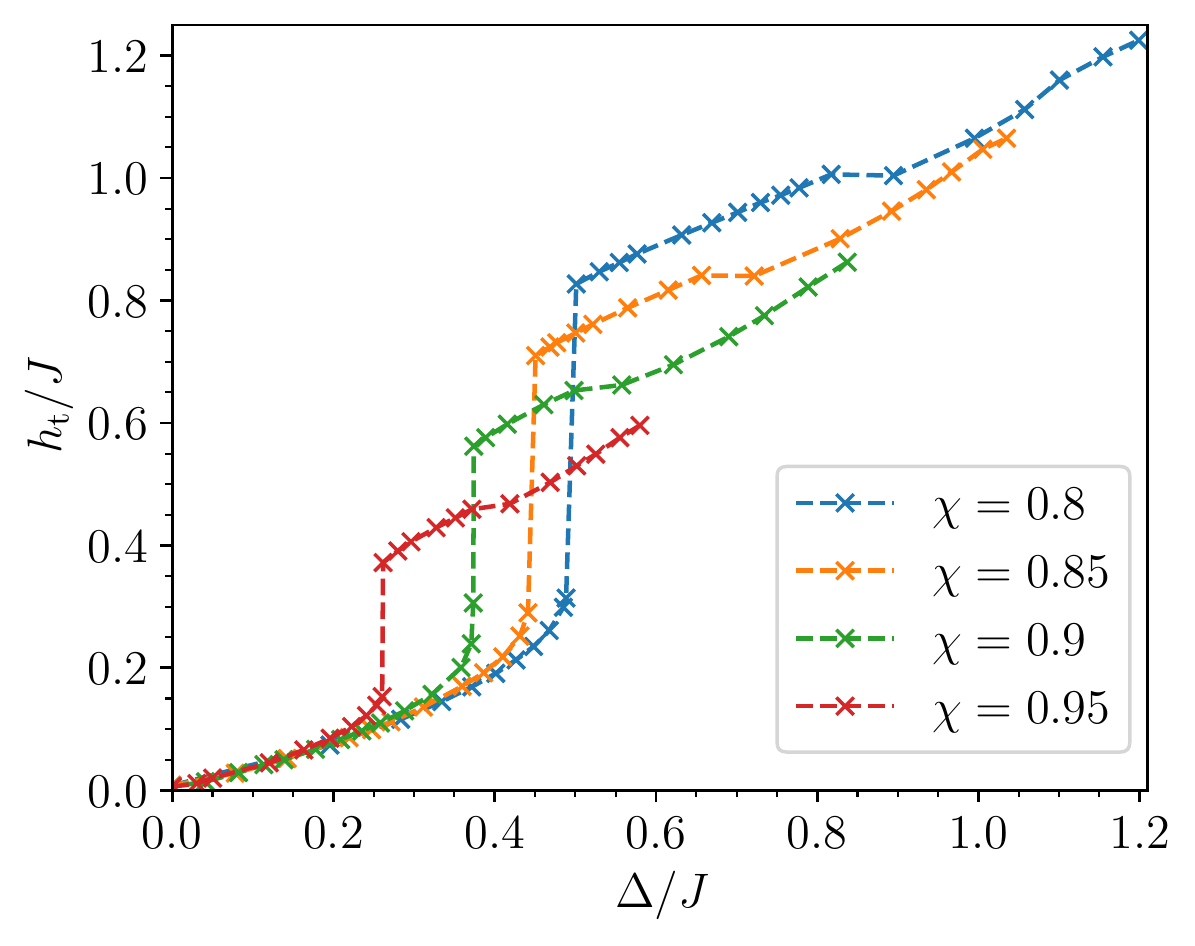}
        \caption{Threshold magnetic field $h_\text{t}$ vs.\ temperature for various anisotropies. 
The data is obtained for a system size $L = 200$. The jump occurs for $\chi=0.95$ at $T=0.6215J$ with
$T_\text{N}=0.652J$, for $\chi=0.9$ at $T=0.6775J$ with $T_\text{N}=0.704J$, for $\chi=0.85$ at $T=0.7065J$ with
$T_\text{N}=0.725J$, and for $\chi=0.8$ at $T=0.7225J$ with
$T_\text{N}=0.732J$.
}
        \label{fig:Schwellfeld-Tfinite}
\end{figure}

Next, it is important to track the threshold fields for various temperatures
to learn how far they can be reduced by increasing the temperature
up to $T_\text{N}$. Since the greatest effects occur close to the 
N\'eel temperatures we do not plot the threshold fields as function of
$T$, but as function of the effective spin gap $\Delta$ in  
Fig.\ \ref{fig:Schwellfeld-Tfinite}.
We stress that there is a monotonic one-to-one mapping between the 
temperature and $\Delta$, see Fig.\ \ref{fig:gaps-Tfinite}(b).
Clearly, the data supports the finding that the threshold field $h_\text{t}$ 
decreases reducing the effective spin gap by increasing the temperature to the 
N\'eel temperature. But the almost quantitative agreement between threshold field
and spin gap we found at zero temperature, see Fig.\ \ref{fig:Schwellfeld},
does not hold anymore. It would have meant that Fig.\ \ref{fig:Schwellfeld-Tfinite}
displayed two straight lines through the origin with identical slope of one.
A truly unexpected feature is the discontinuous jump at an intermediate value
of the spin gap. It appears to be generic since it  occurs for all anisotropies studied.
Below the jump only fairly small fields are required to switch the magnetization.
For application purposes, we conclude that increasing the temperature close to 
$T_\text{N}$ can help significantly to switch the magnetization. Hence, one may 
envisage that writing magnetic data is done at elevated temperatures while the
long-time storage is done at low temperatures where the sublattice magnetization
is considerably more robust.

\begin{figure}[htb]
        \centering
        \includegraphics[width=\columnwidth]{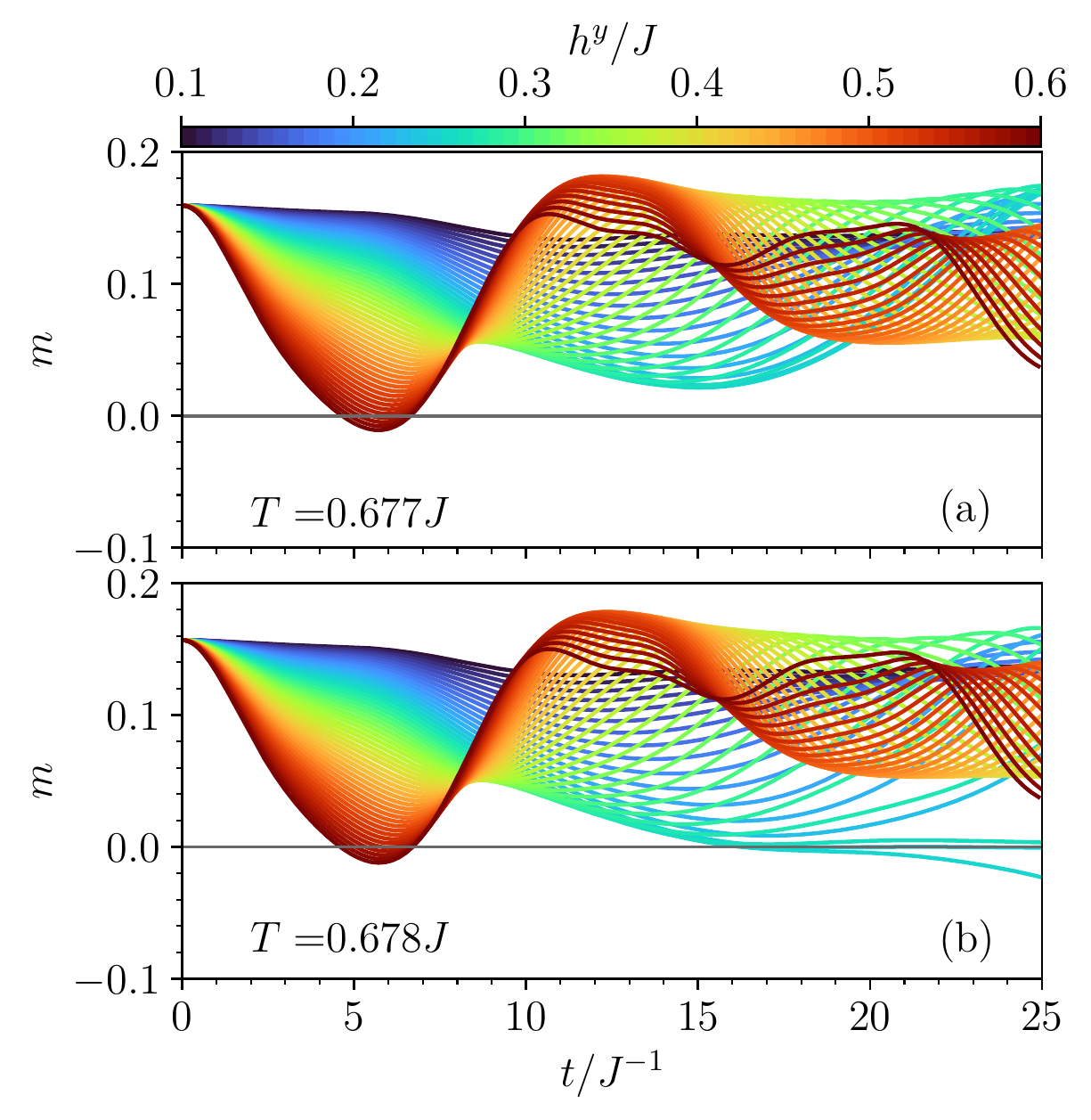}
        \caption{Both panels show the full temporal evolution $m(t)$
				for $\chi=0.9$ for two very close values of the temperature
				of which one is just below (panel (a)) the temperature where the jump of $h_\text{t}$ occurs
				and the other just above (panel (b)). There is a sign change in $m(t)$ in panel (a) 
				at around $t\approx 6J^{-1}$ for larger fields. But note the additional change of sign
				in $m(t)$ in panel (b) at a larger time $t\approx 17J^{-1}$ for lower fields. This is responsible
				for a sudden change of the threshold field.
}
        \label{fig:jump}
\end{figure}

In view of the discontinuity the imminent next question concerns the origin of
this jump. To this end, we show in Fig.\ \ref{fig:jump} the full temporal
evolution of $m(t)$  for two temperatures: one is just below the
jump, i.e., with an effective spin gap $\Delta$ slightly larger than
the value at the jump and the other temperature just above the jump, i.e.,
with an effectve spin gap slightly smaller than the jump value. 
Both panels show successful switching for some magnetic fields.
But for the lower temperature (larger $\Delta$)
the magnetization only switches for large fields at times $t\approx 6\,J^{-1}$. 
For the larger temperature (smaller $\Delta$)
the magnetization also switches for low fields at times $t\approx 17\,J^{-1}$. 
At these times, the lower temperature does not
yet allow for a sign change of the magnetization. Thus, the jump 
of the instant in time at which switching can be detected explains in turn
the jump in the threshold fields.

    \section{Conclusions}
    \label{s:conclusion}
		
Controlling the magnetization of long-range ordered quantum magnets is a key element
in data storage in nanoscale domains. The magnetization orientation in such a
domain serves as a bit. So far, it is realized and employed for ferromagnets. But it is established
that quantum antiferromagnets display important advantages. They do not have
stray fields which oppose close packing of the domains of magnetization, i.e., the bits.
In addition, the generic time scales are shorter by three orders of magnitude in comparison
to generic ferromagnets.

For these reasons, we investigated the switching of the antiferromagnetic sublattice
magnetization in an anisotropic easy-axis Heisenberg antiferromagnet. We aimed at 
the development of a suitable microscopic quantum approach to describe this phenomenon. 
For simplicity, we studied the model on a square lattice. The aim was to go beyond the description
of the magnetizations on the two sublattices by two classical vectors (vector model). 
We intended to base our approach on the quantum model which captures all leading 
quantum and thermal effects. Such a model has to comprise all the magnetic modes, 
i.e., a spin-wave description was required. Addressing all modes allows one to 
deal with dephasing of the modes in the course of switching as well as 
with finite temperature effects. These important effects are missed
otherwise.

But the showstopper of conventional spin-wave theories
is that they only capture the fluctuations around one of the degenerate ground states.
Yet for the purpose of switching from one ground state (up-down on sublattice
A-B) to the other (down-up), this is obviously not sufficient. Hence we resorted 
to the Schwinger boson description which captures all degenerate ordered states
as well as the disordered ones. We use the established mean-field description which
reproduces the result of the usual self-consistent spin-wave theories in equilibrium
based on the Holstein-Primakoff or the Dyson-Maleev representations.

We computed the spin gap within the Schwinger mean-field theory \cite{auerb94} and found results
which agree with the reliable results from other techniques if the mean-field gap
is scaled down by a factor $\approx 1.3$. This is a very satisfying result in view of the
simplicity of the mean-field approach and the low dimensionality of the system.
Similarly, the mean-field temperature $T_{N}$ in the Ising limit agrees with the rigorous result
from Onsager \cite{onsag44} within $30$\%. 

Using the equilibrium expectation values as starting values we computed
solutions of the Heisenberg equations of motions for the expectation values. 
In particular, we calculated the temporal evolution $m(t)$ upon application of
a uniform transverse magnetic field which induces a Larmor precession. The 
magnetization in the isotropic model is rotated without loss of coherence
and for arbitrarily weak fields. In the anisotropic case, however, an activation
energy needs to be overcome. For weak magnetic fields $h$, no switching is possible; only
weak oscillations below the equilibrium magnetization are induced. Above a threshold value
$h_\text{t}$ switching is possible, but the coherence of all the involved magnetic
modes is deteriorating. Thus, for fields just above the threshold only a single
switching is possible while for large control fields several swaps $m_0 \leftrightarrow -m_0$
can be realized. The threshold fields $h_\text{t}$ at zero temperature agree
almost quantitatively with the spin gap. The larger the spin gap the more robust 
the magnetic order is.

Analyzing the time $t_\pi$ needed to perform (or to fail) a swap displays a logarithmic
divergence at the threshold value, both from above and from below. This behavior
coincides precisely with the time needed for a massive particle to overcome an energy barrier.
This means that the antiferromagnetic magnetization disposes of an inertia in its
dynamics as was observed both experimentally and theoretically in a classical vector model
before \cite{kimel09}.

Upon increasing the temperature the equilibrium magnetization and the effective spin gap 
decreases towards the N\'eel temperature where both vanish. Thus, it is not 
surprising that the threshold magnetic field required for switching decreases upon
increasing temperature and vanishes also at $T_\text{N}$.
Our model including the leading quantum and thermal effects confirms this expectation 
quantitatively. The relation between temperature or equivalently the effective spin gap
and the threshold field is monotonic. Unexpectedly, however, we found a discontinuity
in the threshold fields. At an intermediate value of the effective spin gap $\Delta$ 
the required minimum field suddenly decreases by a finite amount. Due to the square root
laws $\Delta \propto \sqrt{T_\text{N} -T}$ this spin gap value corresponds to temperatures which
are close to the transition temperature $T_\text{N}$.
We could trace the origin of the jump to the full temporal evolution of the
magnetization $m(t)$ during the switching. The instant of time where  $m(t)$ changes
sign jumps as well to longer times for lower fields.

What is the implication for experiment? As pointed out above, many parameters
need to be taken into account. If we assume $J=10$ meV and a small anisotropy
$\chi\approx 0.99$ and/or a temperature rather close to the N\'eel temperature,
the threshold field corresponds roughly to $0.05 J = 0.5$ meV which corresponds for
$g=2$ to about 5 Tesla. Note that the spin gaps and thus threshold fields 
are likely to be 30\% lower than the mean-field approach predicts.
This is still a large field, but it is certainly realizable
in a laboratory. Hence, we think that our results provide an interesting and 
quantitative guideline for future experiments. According to our findings
the temperature dependence of the switching  merits close
inspection in particular.

The theoretical outlook comprises a large scope of promising extensions. 
Clearly, the present calculationcan be extended to three dimensions 
and also to many other lattices. Certainly, other
bipartite lattices can be treated in the very same fashion, 
but also frustrated lattices displaying long-range
order such as the triangular lattice \cite{chern09} can be tackled. 
One can also apply the approach to models with anisotropies beyond easy-axis, for instance
with a four-fold rotation symmetry of the magnetizations.
Furthermore, time-dependent
control fields can be considered as well so that a plethora of fundamentally interesting
as well as practically relevant issues such as the influence of
nanostructured confinement, are open for further investigation using
the approach advocated here.

    \begin{acknowledgments} 
      We thank Davide Bossini for drawing our attention to this topic and 
			Christoph Lange for very useful discussions. 
      We gratefully acknowledge financial support by the  German Research Foundation (DFG) in project
			UH 90-14/1 (GSU) and in TRR 160 (AK). Furthermore, KB thanks the Studienstiftung des Deutschen Volkes for
			funding. 
    \end{acknowledgments}


%

    \begin{appendix} 
		 \section{Sublattice magnetization for the isotropic case}
      \label{a:magnetization}

In this appendix, we show how spontaneous symmetry breaking manifests itself in  the Schwinger boson representation 
as Bose-Einstein condensation \cite{auerb94} and how the condensate is related to 
a finite sublattice magnetization, see Eq.\ \eqref{eq:magnet-condensate} in the main text. 
We consider the isotropic antiferromagnetic two-dimensional square lattice at zero temperature 
in the thermodynamic limit. 

In order to capture the spontaneous symmetry breaking we 
initially add an alternating magnetic field $\ha$ in $z$-direction to the system
by means of an alternating Zeeman term \eqref{eq:alternat}. Once the thermodynamic limit is taken,
this external field will be decreased to zero.
The Hamiltonian reads
\begin{equation}\label{eqn:SchwingerH5}
 \mathcal{H} = \mathcal{H}_{0} = \mathcal{H}_{0} - \frac{\ha}{2}\sum_i \lr{\nbi{a} - \nbi{b})},
\end{equation}
where $\mathcal{H}_{0}$ denotes the unperturbed mean-field Hamiltonian \eqref{eqn:SchwingerH3}.
The alternation is no longer manifest because we effectively rotated the spins on every second site
around $S^y$. Since $a$- and $b$-bosons are no longer equivalent in \eqref{eqn:SchwingerH5} the 
two dispersions differ 
\begin{align}
\mathcal{H} &= E_\text{MF} - N\lambda + \nonumber
\\ & \quad 
\sumk\lrg{\omega^\alpha_{\mathbf{k}}\lr{\nbk{\alpha}+\frac{1}{2}} + 
\omega^\beta_{\mathbf{k}}\lr{\nbk{\beta}+\frac{1}{2}}} 
\end{align}
with
\bes
\begin{align}
\omega^\alpha_{\mathbf{k}} &=  \sqrt{\lr{\lambda - {\ha/2}}^{2}
 - (2A\gam)^2} ,
\\
 \omega^\beta_{\mathbf{k}} &=  \sqrt{\lr{\lambda + \ha/2}^{2}- (2A\gam)^2}.
\end{align}    
\ees
Note that for a positive external field, the dispersion of the $\alpha$-boson is lower and hence the
condensation of $\alpha$-bosons is to be expected.

The spontaneous sublattice magnetization results from
\begin{subequations}
\begin{align}
    m_0 &= \lim_{\ha \to 0^+} m(\ha)\qquad \text{with} \\
    m(\ha) &= \lim_{N \to \infty} \frac{1}{2N}\sum_i\lr{\lara{\nbi{a}}-\lara{\nbi{b}}}.
\end{align}    
\end{subequations}
Explicit calculation 
of the average boson number and of the sublattice magnetization yields
\begin{subequations}
\label{eq:SSCG}
\begin{align}
  S + \frac{1}{2} &= \frac{1}{4N} \sumk \lr{\frac{\lambda - \ha/2}{\omega^\alpha_{\mathbf{k}}} 
	+ \frac{\lambda + \ha/2}{\omega^{\beta}_{\mathbf{k}}}}
	,\label{eqn:SSCG1} \\
  m(\ha) &= \frac{1}{4N} \sumk \lr{\frac{\lambda - \ha/2}{\omega^\alpha_{\mathbf{k}}} 
	- \frac{\lambda + \ha/2}{\omega^{\beta}_{\mathbf{k}}}}.
	\label{eqn:SSCG2}
\end{align}
\end{subequations}
Since we expect the condensation of the $\alpha$-bosons, we substitute
 $\lr{\lambda - {\ha/2}}^2 = 4A^2(1+\kappa^2)$
with $\kappa = \tilde f/N$ analogous to what we did in the main text.
This implies 
\begin{subequations}
    \begin{align}
        \omega^\alpha_{\mathbf{k}} &= 2A\sqrt{1+\kappa^2 - \gam^2} , \\
        \omega^{\beta}_{\mathbf{k}} &= 2A \sqrt{1+\kappa^2 + \frac{\lambda h}{2 A^2} - \gam^2} .
    \end{align}
\end{subequations}
The dispersion $\omega^\alpha_{\mathbf{k}}$ becomes gapless in the thermodynamic limit
which is necessary for the condensation of the $\alpha$-bosons.
Transforming the sums in \eqref{eqn:SSCG1} and \eqref{eqn:SSCG2} into integrals 
plus a contribution  at $\mathbf{k}=(0,0)$ and $(\pi,\pi)$
reveals that \emph{no} such extra contribution occurs for the $\beta$-bosons (second fractions
in the brackets in Eqs.\ \eqref{eq:SSCG}) because
at any finite $\ha$, their dispersion does not vanish. Thus, their sum converges
smoothly to the integral contribution. For the $\alpha$-bosons, however, 
we obtain the additional contribution $2/(4\tilde f)=1/(2\tilde f)$ accounting for
the prefactors of the sums in \eqref{eq:SSCG}. The integrals for
the $\alpha$- and the $\beta$-bosons become the same for $\ha\to0$ and thus cancel.
Eventually, we obtain
\be
m_0= 1/(2\tilde f).
\ee
This relates the spontaneously occurring sublattice magnetization directly to the
condensate fraction, here of the $\alpha$-bosons.

In order to arrive at Eq.\ \eqref{eq:magnet-condensate} we recall that there the choice $\kappa=f/N$ 
referred to the \emph{disordered} case where \emph{both} bosons, $\alpha$ and $\beta$, condense
yielding a condensate contribution $\propto 1/f$. Thus, if only \emph{one} kind of boson
condenses, it must contribute double the amount, hence $\tilde f = f/2$ due to the
inverse proportionality. Hence $m_0=1/f$ ensues.

For an explicit calculation, one can subtract  the two equations 
 \eqref{eqn:SSCG1} and \eqref{eqn:SSCG2}
\begin{equation}
        S + \frac{1}{2} - m_0 = \frac{1}{8\pi^2} \int_\text{BZ} dk^2 \frac{1}{\sqrt{1-\gam^2}}.
\end{equation}
This eliminates the macroscopic contributions from  $\mathbf{k}=(0,0)$ and $(\pi,\pi)$.
Then we set $\ha=0$ to reach the above equation. Solving for $m_0$ yields
\bes
\begin{align}
\label{eq:m0}
        m_0 &= S + \frac{1}{2} -\frac{1}{8\pi^2} \int_\text{BZ} d k^2 \frac{1}{\sqrt{1-\gam^2}} \\
				&= S - 0.1966 , \label{eqn:Magn}
\end{align}
\ees
where we evaluated the integral numerically for the last line. Comparing
the first line \eqref{eq:m0} with half the equation \eqref{eq:selfcon2b}
one arrives at $m_0=1/f$ as stated in Eq.\ \eqref{eq:magnet-condensate}.
This underlines the above argument that $\tilde f=f/2$ holds in the 
comparison of the long-range ordered ground state with the disordered ground state.
In summary, Eq.\ \eqref{eq:m0-explicit} is derived.
		
\section{Approximating the thermodynamic limit for the isotropic case}
      \label{a:h-aux} 

A finite system is used to approximate an infinitely large 
lattice in order to obtain a numerically feasible set of differential equations.
However, unlike the infinite or anisotropic case, finite two-dimensional isotropic spin lattices,
i.e., for $\chi=1$, do not display magnetic order.
Therefore, an infinitesimal magnetic field $\ha$ in $z$-direction is added, which enforces an initial magnetic order
to make the finite isotropic system have the same magnetization as obtained in \eqref{eqn:Magn},
which is $m_{0,\text{iso}} = 0.3034$ for $S=\tfrac{1}{2}$.

\begin{figure}[htb]
    \centering
    \includegraphics[width=\columnwidth]{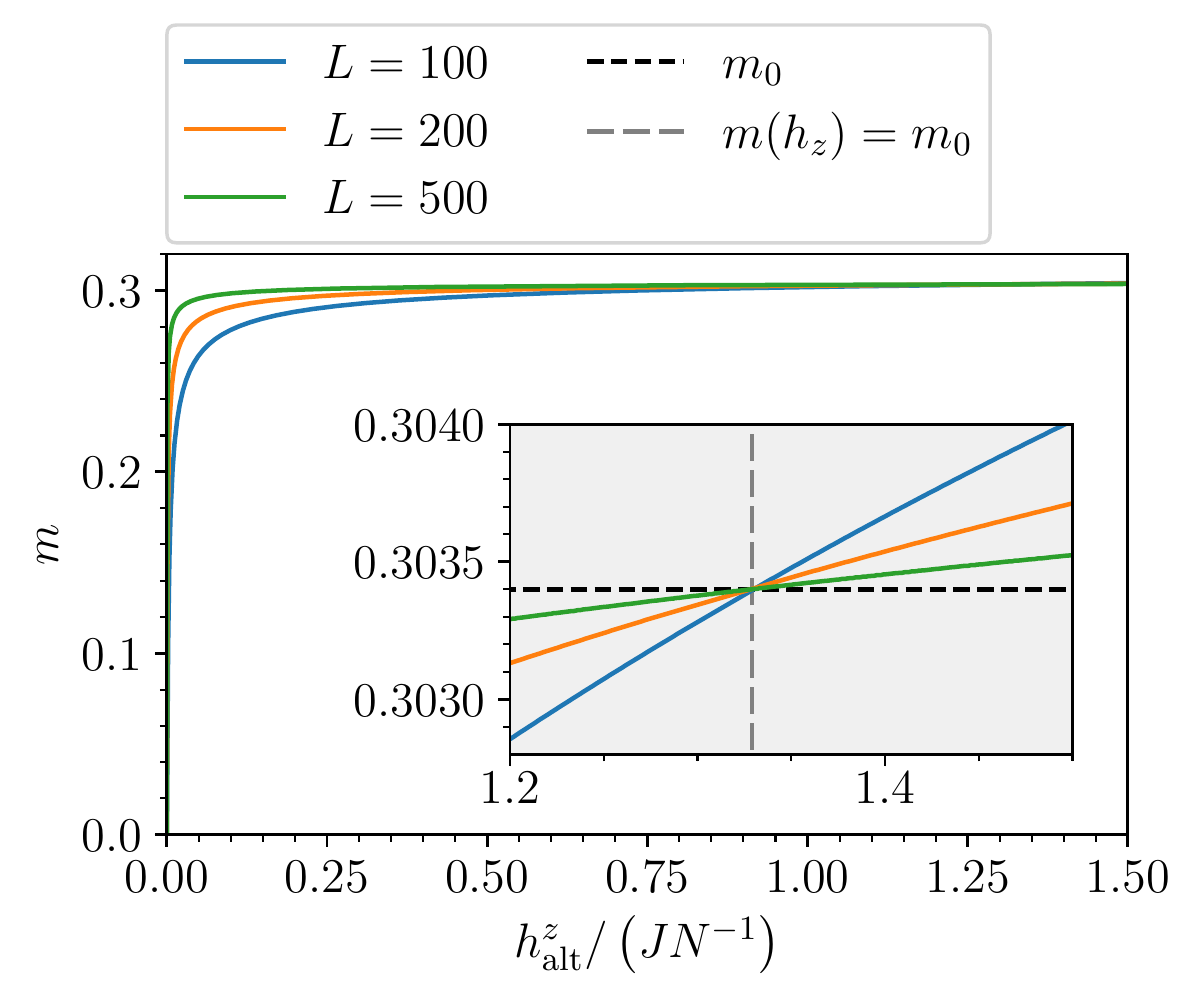}
    \caption{The initial magnetization $m$ of the isotropic system, i.e., $\chi=1$, as a 
    function of a small field $\ha$ for different system sizes.
    Magnetic order is already obtained for small fields, recall $N=L^2$.
		The desired initial magnetization $m_{0,\text{iso}} = 0.3034$ is reached for $\ha = 1.329 JN^{-1}$, 
    as shown in the inset. Around this value, the curves for the various system sizes intersect.
		}
    \label{fig:MagneInfiFeld}
\end{figure}

In order to determine a suitable value of the magnetic field, the initial magnetization $m(\ha)$ 
of the isotropic system 
for various system sizes and spin $S=\tfrac{1}{2}$ is plotted against the scaled field $\ha$ in 
Fig.\ \ref{fig:MagneInfiFeld}.
All curves demonstrate that a small field scaling $\propto \frac{1}{N}$ is already sufficient to generate 
the magnetic order.
The larger the system, the more the magnetization curve converges to the discontinuous curve 
of an infinitely large system, where the magnetization persists even for $\ha\to 0^+$.
The inset shows that the sublattice magnetization $m(\ha)$ corresponds to the 
desired value $m_{0,\text{iso}} = 0.3034$ for $\ha = 1.329\,JN^{-1}$.
We find it remarkable and reassuring that in the close vicinity of this value, the curves
for various $L$ almost intersect. They do not intersect precisely in one point, but
in a very narrow region. Still, this corroborates our way to approximate the thermodynamic limit
by finite clusters.

\section{Zero temperature equations for the anisotropic case}
      \label{a:magnetization-anisotropic} 

\begin{figure}[htb]
    \centering
    \includegraphics[width=0.8\columnwidth]{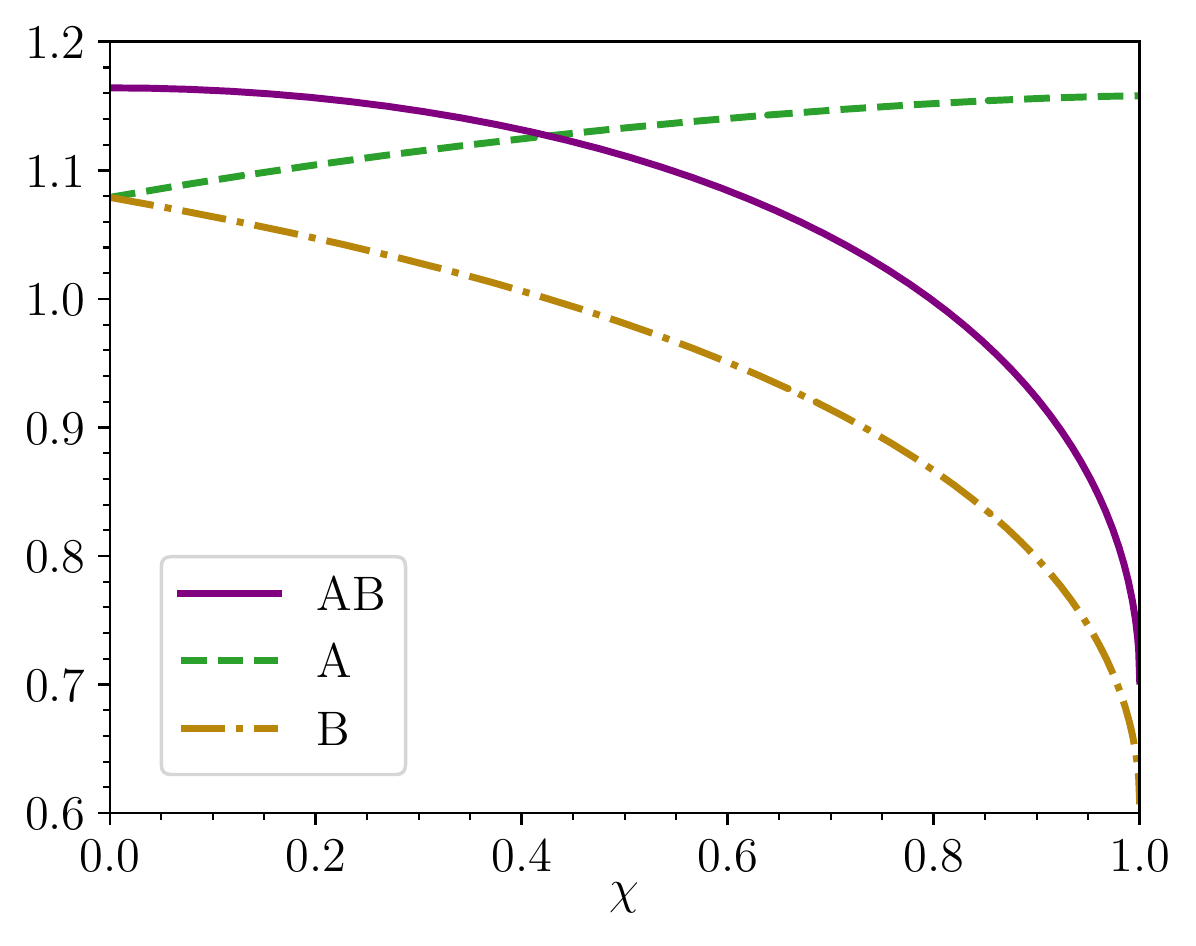}
   \caption{Expectation values $A$ and $B$ and their product in equilibrium
	   plotted as a function of the anisotropy $\chi$. These values enter in the determination of 
		the spin gap in Fig.\ \ref{fig:AnisoGap} and of the magnetization in Fig.\ \ref{fig:magnetization}.}
    \label{fig:AB}
\end{figure}

For the condensation of the $\alpha$-bosons, the dispersion $\omega^{-}$ should become gapless
for an infinitely large system at zero temperature, which is why 
\be
\label{eq:lambda}
\lambda^2 = C_-^2(1+\kappa^2)
\ee
 is chosen with
 $\kappa = \frac{\tilde f}{N}$. We use $\tilde f$ because only one of the two boson flavors
is to condense. As before, this yields the dispersions
\begin{subequations}
\begin{align}
    \omega^{-}_\mathbf{k} &= C_-\sqrt{1+\kappa^2-\gam^2}\label{eqn:dis1} ,\\
    \omega^{+}_\mathbf{k} &= \sqrt{C_-^2(1+\kappa^2)-C_+^2\gam^2}
		\label{eqn:dis2} .
\end{align}
\end{subequations}
It is important to note that only $\omega^{+}_\mathbf{k}$ describes the true
spin-wave spectrum for $T=0$ because the operator of the condensed 
boson can be replaced $\alpha^{(\dag)}_\mathbf{k} \to \sqrt{N m_0/2}$.
In the thermodynamic limit $N\to \infty$, the sums in \eqref{eqn:SCEQ} become integrals 
as before plus the contributions from $\mathbf{k}=(0,0)$ and $(\pi,\pi)$
\begin{subequations}
    \begin{align}
        2S &= \frac{1}{8\pi^2}\int_{\text{BZ}}\!\!\!dk^2\Big[\frac{1}{\sqrt{1-\gam^2}}
				+\frac{C_-}{\sqrt{C_-^2 - C_+^2 \gam^2}}\Big] + \frac{1}{\tilde f}-1,\label{eqn:AniSc1}\\
        A &= \frac{1}{8\pi^2}\int_{\text{BZ}}\!\!\!dk^2\gam^2\Big[\frac{1}{\sqrt{1-\gam^2}}
				+\frac{C_+}{\sqrt{C_-^2 - C_+^2 \gam^2}}\Big] + \frac{1}{\tilde f}, \label{eqn:AniSc2}\\
        B &= \frac{1}{8\pi^2}\int_{\text{BZ}}\!\!\!dk^2\gam^2\Big[\frac{1}{\sqrt{1-\gam^2}}
				-\frac{C_+}{\sqrt{C_-^2 - C_+^2 \gam^2}}\Big] + \frac{1}{\tilde f} .\label{eqn:AniSc3}
    \end{align}
\end{subequations}
Finally, \eqref{eqn:AniSc2} and \eqref{eqn:AniSc3} are 
each subtracted from \eqref{eqn:AniSc1}, yielding the two equations
\begin{subequations}
 \begin{align}
        2S+1 - A &= \frac{1}{8\pi^2}\int_{\text{BZ}}\!\!\!dk^2\Big[\frac{1-\gam^2}{\sqrt{1-\gam^2}}
				+\frac{C_- - C_+ \gam^2}{\sqrt{C_-^2 - C_+^2 \gam^2}}\Big] ,\\
        2S+1 - B &= \frac{1}{8\pi^2}\int_{\text{BZ}}\!\!\!dk^2\Big[\frac{1-\gam^2}{\sqrt{1-\gam^2}}
				+\frac{C_- + C_+ \gam^2}{\sqrt{C_-^2 - C_+^2 \gam^2}}\Big] .
    \end{align}
\end{subequations}
These equations allow us to determine $A$ and $B$. The results are shown in Fig.\ \ref{fig:AB}. 
The Lagrange multiplier $\lambda$ is
implicitly fixed by the condition that the $\alpha$-bosons condense and their dispersion is 
massless.

To determine the sublattice magnetization from Eq.\ \eqref{eq:magnet-aniso-sum} we consider its 
thermodynamic limit $N\to\infty$
\be
m_0 = \frac{1}{16\pi^2}\int_{\text{BZ}}\!\!\!dk^2\Big[\frac{1}{\sqrt{1-\gam^2}}
				-\frac{C_-}{\sqrt{C_-^2 - C_+^2 \gam^2}}\Big] + \frac{1}{2\tilde f}.
\ee
Subtracting this equation from half the Eq.\ \eqref{eqn:AniSc1} yields
\be
S+\frac{1}{2}-m_0 = \frac{1}{8\pi^2}\int_{\text{BZ}}\!\!\!dk^2 \frac{C_-}{\sqrt{C_-^2 - C_+^2 \gam^2}} 
\ee
eliminating the condensate contribution $\propto 1/\tilde f$.
Solving the last equation for $m_0$ yields Eq.\ \eqref{eq:magnet-aniso}.

\section{Overcoming a potential barrier}
      \label{a:barrier} 
			
\begin{figure}[htb]
    \includegraphics[width=0.8\columnwidth]{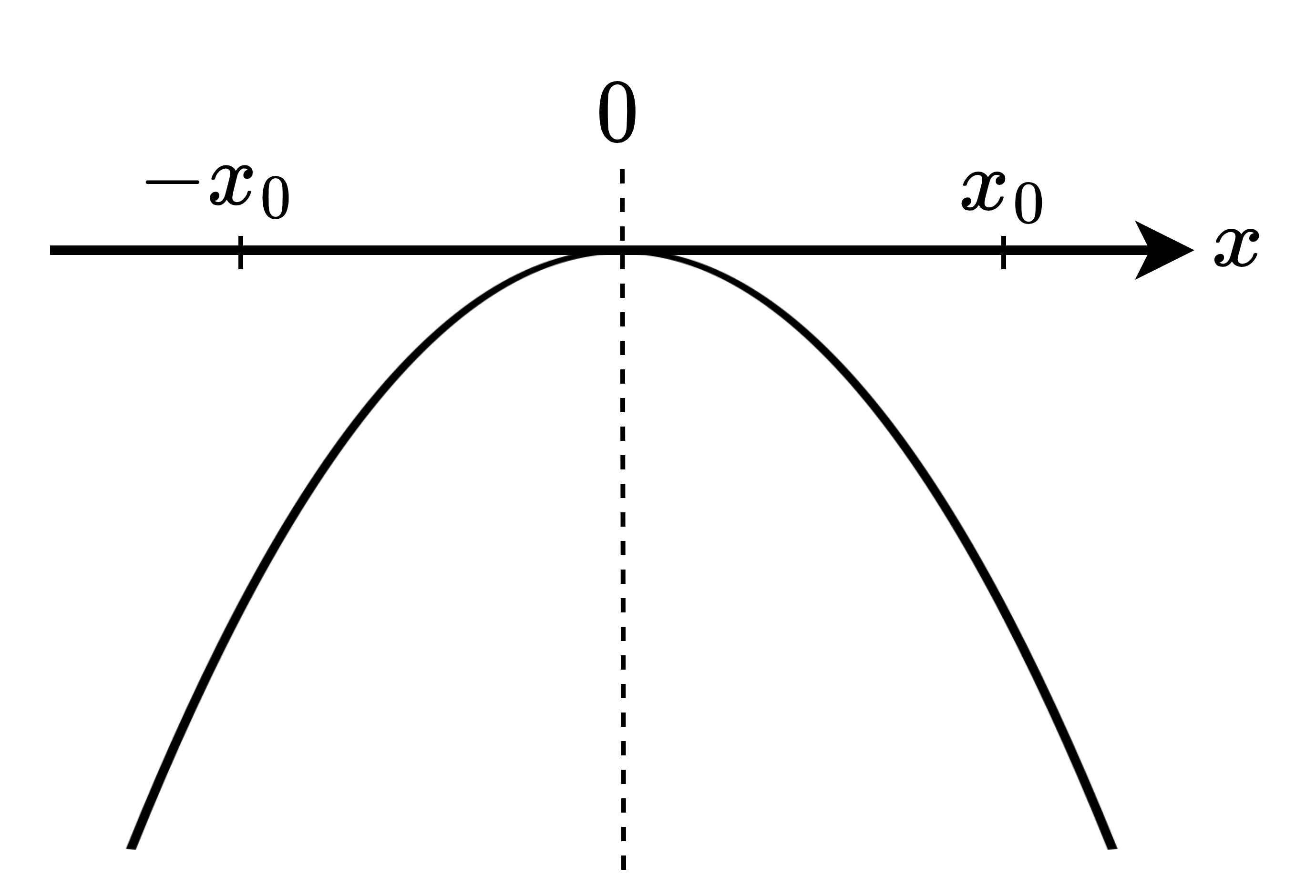}
      \caption{The considered potential barrier ${V(x)=-\frac{1}{2}\gamma^2x^2}$. 
      The points $\pm x_0$ are used to indicate whether the barrier has been overcome or not.}
      \label{fig:PotBar}
\end{figure}

Here we motivate by a classical example why a logarithmic singularity is to 
be expected in Fig.\ \ref{fig:Hthres2} and why this is an indication of inertia.
For this purpose a potential ${V(x)=-\frac{1}{2}\gamma^2x^2}$ is considered as shown 
in Fig.~\ref{fig:PotBar}.
We want to know how long it takes a mass to move over this 
potential barrier. 
For the sake of simplicity, the mass is set to $m=1$.
The Hamilton function of the system reads
\begin{equation}
    \mathcal{H} = \frac{p^2}{2} - \frac{1}{2}\gamma^2x^2.
\end{equation}
We highlight that the existence of the kinetic energy
reflects the existence of inertia.
The resulting equation of motion reads
\begin{equation}
    \ddot{x} - \gamma^2 x = 0 ,
\end{equation}
and has  the general solution
\begin{equation}
    x(t) = F \cosh(\gamma t) + G\sinh(\gamma t) .
\end{equation}

The initial conditions are $x(0) = x_0 >0$ and $\dot{x}(0) = - v_0 <0$.
Thus, we find $F=x_0$ and $G=-v_0/\gamma$. The total energy of the system takes the value
\begin{equation}
    2E = v_0^2 - \gamma^2x_0^2 = (\gamma G)^2 - (\gamma F)^2 .
\end{equation}
The marginal case is given for $E=0$ because in this case the kinetic 
energy disappears exactly when the potential maximum is reached.
This corresponds to $F = -G$ implying
\begin{equation}
\label{eq:class-solution}
    x(t) = x_0 \lr{\cosh(\tau) - \sinh(\tau)} = x_0 e^{-\tau} 
\end{equation}
with $\tau \coloneqq \gamma t$.
Note that the solution $F=G$ is discarded because of the restrictions $F>0$ and $G<0$.
The maximum at $x=0$ is thus reached exponentially slowly.
Next, we consider a small deviation ${G = -x_0(1+2\delta)}$ from the marginal case.
For $\delta > 0$, the total energy is positive, and therefore the potential barrier can be passed. 
We calculate the necessary time when the point $-x_0$ is reached
\begin{subequations}
\begin{align}
    -x_0 & = x_0 e^{-t} - \delta x_0 \lr{e^{\tau} - e^{-\tau}} \\
    \Leftrightarrow \quad 0 &= \delta (y^2 - 1) - y - 1 \quad \text{with} \ y\coloneqq e^\tau \\
    \Leftrightarrow \quad y &= \frac{1}{2\delta} + \sqrt{\frac{1}{4\delta^2} + \frac{1 + \delta}{\delta}} \\
     &= 1+\frac{1}{\delta} + \mathcal{O}(\delta^2) \, .
\end{align}
\end{subequations}
The other solution of the quadratic equation is negative and therefore not a physical solution.
The position $-x_0$ is reached at the time
\begin{equation}\label{eqn:res1}
    \gamma t = \tau = -\ln|\delta| + \delta + \mathcal{O}(\delta^2) .
\end{equation}

For $\delta < 0$, the total energy is negative, and therefore the potential barrier cannot be passed;
the passage fails. 
In this case, we calculate the time it takes to get back to the point $x_0$
\begin{subequations}
    \begin{align}
    x_0 &= x_0 e^{-\tau} + |\delta|x_0 \lr{e^{\tau} - e^{-\tau}} \\
        \Leftrightarrow \quad  0 &= y - 1 - |\delta|\lr{y^2 -1} \\
        \Leftrightarrow \quad y &= \frac{1}{2|\delta|} + \sqrt{\frac{1}{4|\delta|^2} - \frac{1-|\delta|}{\delta}}\\ 
				& = \frac{1}{|\delta|} - 1 + \mathcal{O}(\delta^2) .
    \end{align}
\end{subequations}
Again, the other solution of the quadratic equation is not physically significant.
So the time for reaching $x_0$ again is 
\begin{equation}
    \gamma t = \tau =  -\ln|\delta| + \delta + \mathcal{O}(\delta^2).
\end{equation}
Remarkably, the same result ensues for succeeding to pass the barrier as for failing to pass it.
In both cases, a logarithmic  divergence of the time occurs just as we observed for
switching the sublattice magnetization in Figs.\ \ref{fig:scan-magnet-field} and 
\ref{fig:Hthres2}. Note that the fitted prefactors in Eqs.\ \eqref{eq:succeeded} and 
\eqref{eq:failed} are very close
to each other as one expects from the classical calculation presented above.
Note that this symmetry is perfectly reflected by the curves in Fig.\ \ref{fig:scan-magnet-field}
for magnetic fields just above and just below the threshold value
which are mirror images of one another in the vicinity of the first
extremum for $t>0$.

In addition, we observe in the solution \eqref{eq:class-solution} as well
as in the curves in Fig.\ \ref{fig:scan-magnet-field} for the cases in the
vicinity of the threshold that most of the time is spent around the energy maximum $x\approx0$
or $m\approx0$, respectively. This underlines that we are dealing with a process
governed by inertia: What matters is the maximum \emph{energy} that the switching term can 
provide. In contrast, a process without inertia, governed by friction, would spend most of the
time close to the point where the maximum \emph{force} is required. This is certainly not the
case for $m\approx 0$, but rather would be around $m\approx \pm m_0/2$.
Thus, the curves in Fig.\ \ref{fig:scan-magnet-field} underline the conclusion that
the magnetization dynamics in quantum antiferromagnets is governed by inertia.

\end{appendix}	

\end{document}